\documentclass[11pt]{article}
\pdfoutput=1

\newlength{\vshift}
\newlength{\hshift}
\usepackage{amssymb,amsopn}
\usepackage{slashed}
\usepackage[utf8]{inputenc}
\usepackage{amsmath}
\usepackage{graphicx}
\usepackage{color}
\renewcommand{\theequation}{\thesection.\arabic{equation}}
\newcommand{\initiate}{\setcounter{equation}{0}}
\definecolor{zelena}{rgb}{0.0, 0.488, 0.0}

\def\d{\textrm{d}}
\def\ds{\stackrel{\star}{,}}

\def\nn{\nonumber}
\def\be{\begin{equation}}             \def\ee{\end{equation}}
\def\ba#1{\begin{array}{#1}}          \def\ea{\end{array}}
\def\bea{\begin{eqnarray} }           \def\eea{\end{eqnarray} }
\def\beann{\begin{eqnarray*} }        \def\eeann{\end{eqnarray*} }
\def\beal{\begin{eqalign}}            \def\eeal{\end{eqalign}}
             
\def\bsubeq{\begin{subequations}}     \def\esubeq{\end{subequations}}
\def\bitem{\begin{itemize}}           \def\eitem{\end{itemize}}

\begin{document}

\begin{titlepage}

\begin{center}

{\large{\bf Noncommutative scalar field in the non-extremal \\ 
Reissner–Nordstr\" om background: QNM spectrum}}

\vspace*{1.5cm}

{{\bf Marija Dimitrijevi\' c \' Ciri\'c${}^{1}$, 
Nikola Konjik${}^{1}$ and Andjelo Samsarov${}^{2}$}}

\vspace*{1cm}

\noindent  ${}^1$ {\it Faculty of Physics, University of Belgrade}\\ {\it Studentski trg 12,
11000 Beograd, Serbia}\\

\noindent  ${}^2$ {\it Rudjer Bo\v skovi\' c Institute, Theoretical Physics Division\\
Bijeni\v cka 54, 10002 Zagreb, Croatia}\\

\end{center}

\vspace*{2cm}

\begin{abstract}

\end{abstract}
\vspace*{1cm}

In our previous work \cite{Ciric:2017rnf} we constructed a model of a noncommutative, charged and
massive scalar field
based on the angular twist. Then we used this model to analyze the motion of the scalar field in the
Reissner–Nordstr\" om black hole background. In particular, we determined the QNM spectrum
analytically in the near-extremal limit. To broaden our analysis, in this paper we apply a well
defined numerical method, the continued fraction method and calculate the QNM spectrum for a
non-extremal Reissner–Nordstr\" om black hole. To check the validity of our analytic calculations,
we compare results of the continued fraction method in the near extremal limit with the
analytic results obtained in the previous paper. We find that the results are in good agreement. For
completeness, we also study the QNM spectrum in the WKB approximation.

\vspace*{1cm}

{\bf Keywords:} {NC scalar quasinormal modes, RN black hole, WKB, continued fraction}

\vspace*{1cm}
\quad\scriptsize{eMail:
dmarija@ipb.ac.rs, konjik@ipb.ac.rs, asamsarov@irb.hr}
\vfill

\end{titlepage}

\tableofcontents

\newpage

\section{Introduction}

The study of black hole perturbations  has a long history dating back to the work of Regge and
Wheeler \cite{rg}  and Vishveshwara \cite{vish}. The initial impetus came from an attempt to analyze
the issue of stability of a black hole. Later on, the study was extended  to include  various types
of perturbations in almost all conceivable backgrounds.

With a recent discovery of gravitational waves \cite{ligo} this line of  investigation became ever more important. After being perturbed, black holes  return back to their equilibrium by going  through a ring down phase, whose most dominant stage is characterized by the long lasting damped oscillations dubbed as quasinormal modes (QNMs) \cite{press}. This phase
is characterized by a discrete set of complex frequencies, called QNM frequencies.
While the real part of the frequency corresponds to the actual frequency of the wave dynamics, the
imaginary part represents the damping factor. For an overview of the subject,
we refer the reader to some  excellent reviews  \cite{nollert-review,Kokkotas:1999bd,cardosoreview,KonopReview}.

Several methods have been devised to determine the quasinormal modes of a black hole. They range from purely analytic methods \cite{Cardoso:2001hn,Cardoso:2001bb} through various semiclassical ones \cite{schutzwill,iyer,mashoon,chandra} and  all down to the purely numerical methods \cite{numerical}.

A way to describe some effects of quantum gravity is to introduce a noncommutative deformation of
space-time. The main idea of noncommutative (NC) geometry is that a
space-time as we perceive it might possibly be distorted at some relatively high energy
scale, which we label as the scale of noncommutativity $l_{NC}$. The scale of noncommutativity
itself may be anything between the TeV scale and the Planck scale. Moreover, if  the space-time is
indeed modified at the NC scale $l_{NC}$, 
then this distortion should  somehow be visible in the QNM spectrum of black holes. In other
words, with the current capacities as well as  with the new possibilities that have opened up by the
detection of gravitational waves, a direct comparison of the theoretical results with the
observation may signal a presence of a new physics, i.e. may hint toward a  presence of an
underlying NC space-time  structure above some energy threshold. 

First results on NC QNM spectrum were published in \cite{Gupta:2015uga, Gupta:2017lwk}.
There, the authors analyze the NC scalar QNM spectrum of the three dimensional BTZ black hole using
a direct integration of the equation of motion for the NC scalar field in the BTZ black hole
background. 

In our recent paper \cite{Ciric:2017rnf} we have investigated a model of NC scalar
and gauge fields coupled to a classical background of the Reissner–Nordstr\" om (RN) black hole.
This model resulted in a master equation governing a behavior of the scalar perturbations in
the presence of a noncommutative structure of space-time. The master equation was then
analyzed analytically. However, the analytic treatment is only possible for a highly restricted
range of parameters corresponding to the near extremal case of the RN black hole. In order to
overcome the shortcomings of the analytic treatment, in the present work we analyze the same master
equation by a well defined numerical method, the continued fraction method. This method enables us
to find solutions for QNM frequencies for a more general set of system parameters. Of course, the
results we obtain may also be confronted with the results obtained in \cite{Ciric:2017rnf} 
after specializing them to the near extremal case. Since the analytic treatment of the reference
\cite{Ciric:2017rnf} was carried out under this rather restrictive assumption, the treatment
conducted here could also serve to test the validity of this assumption as well as  the  general
accuracy of our previous analytic treatment.

To make the paper self consistent, in the next section we repeat some of the results from our
previous paper \cite{Ciric:2017rnf}. In particular, we describe the NC deformation by the angular
twist and its consequences on a massive, charged scalar field propagating in the fixed (undeformed)
RN background. Finally, we derive the NC master equation for the scalar field. This equation is our
starting point in Section 3. There we investigate the QNM spectrum of the massive, charged scalar
field by the WKB method. Our analysis is analytic and it is therefore limited by an approximation
$l+1 \ll qQ$, where $l$ is the orbital momentum of the scalar perturbation, $Q$ is the black
hole charge and $q$ is the charge of the scalar perturbation. After this warm-up, in Section 4 we
attack the problem of finding the QNM spectrum with the well defined numerical method, the continued
fraction method. In Section 5 we discuss our results and give some final remarks. Details of a few
cumbersome calculations are given in the Appendix.

\section{NC scalar field in the RN background}

A noncommutative deformation of space-time can be introduced in different ways \cite{NCbooks}. We
follow the twist approach \cite{NCbookMi} and deform the Poincar\'e algebra to a twisted Poincar\'e
algebra. The algebra relations remain the same, while the comultiplication changes. This change is
relevant for multiparticle states \cite{Napulj2018}. One of the advantages of the twist deformation
is that it induces a deformed differential calculus in a well defined way. In particular, it
introduces a deformed product of functions, the $\star$-product.

Our goal is to study the motion of a massive, charged scalar field in the geometry of the RN black hole described by the metric
\begin{equation}
{\rm d}s^2 = (1-\frac{2MG}{r}+\frac{Q^2G}{r^2}){\rm d}t^2 - \frac{{\rm
d}r^2}{1-\frac{2MG}{r}+\frac{Q^2G}{r^2}} - r^2({\rm d}\theta^2 + \sin^2\theta{\rm d}\varphi^2)
. \label{dsRN}
\end{equation}
Here $M$ is the mass of the RN black hole, while $Q$ is the charge of the RN black hole. This problem requires a model of NC gravity coupled with a NC scalar and a NC electromagnetic field. A NC deformation of gravity is not easy to construct, see \cite{EPL2017} and references therein. Therefore, we follow a semiclassical approach based on the assumption that the geometry
(gravitational field) is classical (it is not deformed by noncommutativity), while the scalar field
propagating in the RN background "feels" the effects of space-time noncommutativity. In order to
realize this approach, we choose a Killing twist. The twist is given by
\begin{equation}
\mathcal{F} = e^{-\frac{ia}{2} (\partial_t\otimes\partial_\varphi -
\partial_\varphi\otimes\partial_t)}.
\label{AngTwist0Phi}
\end{equation}
The NC deformation is controlled by the small deformation parameter $a\approx l_{NC}$. Vector fields
$X_{1}=\partial _t$,
$X_{2}= \partial_\varphi$ are commuting vector fields,
$[X_{1},X_{2}]=0$, therefore the twist (\ref{AngTwist0Phi}) is an Abelian twist \cite{PL09}. We call (\ref{AngTwist0Phi}) an "angular twist" because the vector field $X_2 =
\partial_\varphi$ is a generator of rotations around $z$-axis. The vector fields $X_1$ and $X_2$ are two Killing vectors for the metric
(\ref{dsRN}) and that is why the twist (\ref{AngTwist0Phi}) is called a Killing twist. In
particular, the twist
(\ref{AngTwist0Phi}) does not act on the RN metric and it does not act on functions of the RN
metric.
In this way we ensure that the geometry remains undeformed.

The twist (\ref{AngTwist0Phi}) defines the $\star$-product of functions (fields):
\begin{eqnarray}
f\star g &=&  \mu \{ e^{\frac{ia}{2} (\partial_t\otimes \partial_\varphi - \partial_\varphi\otimes
\partial_t)}
f\otimes g \}\nn\\
&=& fg + \frac{ia}{2}
(\partial_t f(\partial_\varphi g) - \partial_t g(\partial_\varphi f)) + 
\mathcal{O}(a^2) .\label{fStarg0Phi}
\end{eqnarray}
Using this product, one can write an action for a massive, charged scalar field in the fixed RN background as
\begin{equation}
S_\phi = \int \d ^4x \, \sqrt{-g}\star\Big( g^{\mu\nu}\star D_{\mu}\hat{\phi}^+ \star
D_{\nu}\hat{\phi} - \mu^2\hat{\phi}^+ \star\hat{\phi}\Big) . \label{SPhi} 
\end{equation}
The scalar field $\hat{\phi}$ is a
complex charged
scalar field with mass $\mu$ and charge $q$. It transforms in the fundamental representation of the
noncommutative $U(1)_\star$ gauge transformations. Therefore its covariant
derivative is defined as
\begin{equation}
D_\mu\hat{\phi} = \partial_\mu\hat{\phi} - i \hat{A}_\mu\star \hat{\phi} \label{DPhi} , \nn
\end{equation}
with NC $U(1)_\star$ gauge field $\hat{A}_\mu$. Note that $\star$-products in $\sqrt{-g}\star
g^{\alpha\beta}\star g^{\mu\nu}$ can all be removed since the twist (\ref{AngTwist0Phi}) does not
act on the metric tensor (\ref{dsRN}).

One can check that the action (\ref{SPhi}) is invariant under the infinitesimal
$U(1)_\star$
gauge transformations defined in the following way:
\begin{eqnarray}
\delta^\star \hat{\phi} &=& i\hat{\Lambda} \star \hat{\phi}, \nn\\
\delta^\star \hat{A}_\mu &=& \partial_\mu\hat{\Lambda} + i[\hat{\Lambda} \ds \hat{A}_\mu],
\label{NCGaugeTransf}\\
\delta^\star g_{\mu\nu} &=& 0,\nn
\end{eqnarray}
with the NC gauge parameter $\hat{\Lambda}$.

Our approach is perturbative and we have to expand the action (\ref{SPhi}) up to first order in the deformation parameter $a$. To do that we expand the $\star$-products in (\ref{SPhi}) and use the Seiberg-Witten (SW) map. SW map enables to express NC variables as functions of the corresponding commutative variables. In this
way, the problem of charge quantization in $U(1)_\star$ gauge theory does not exist. In the case of
NC Yang-Mills theories, SW map guarantees that the number of degrees of freedom in the NC
theory is the same as in the corresponding commutative theory. That is, no new degrees of freedom
are introduced.

Using the SW-map NC fields can be expressed as function of corresponding commutative fields
and can be expanded in orders of the deformation parameter $a$. Expansions for an arbitrary Abelian
twist deformation are known to all orders \cite{PLSWGeneral}. Applying these results to the twist
(\ref{AngTwist0Phi}), 
expansions of fields up to first
order in the deformation parameter $a$ follow. They are given by:
\begin{eqnarray}
\hat{\phi} &=& \phi -\frac{1}{4}\theta^{\rho\sigma}A_\rho(\partial_\sigma\phi + D_\sigma
\phi), \label{HatPhi}\\
\hat{A}_\mu &=& A_\mu -\frac{1}{2}\theta^{\rho\sigma}A_\rho(\partial_\sigma A_{\mu} +
F_{\sigma\mu}). \label{HatA}
\end{eqnarray}
The $U(1)$ covariant derivative of $\phi$ is defined as $D_\mu \phi = (\partial_\mu - i A_\mu)
\phi$ and $F_{\mu \nu} = \partial_\mu A_{\nu} - \partial_\nu A_{\mu}$ is the electromagnetic field
tensor. To have a more compact equations and keep the track of the NC deformation, we introduce the
$4\times 4$ antisymmetric matrix $\theta^{\alpha\beta}$ with the only non-zero components
$\theta^{t\varphi} = -\theta^{\varphi t} =a$. It is also important to note that the coupling
constant $q$
between fields $\phi$ and $A_\mu$, the charge of $\phi$, is included into $A_\mu$, namely $A_\mu =
qA_\mu$. Using the SW-map solutions and expanding the $\star$-products in (\ref{SPhi}) we
find the action up to first order in the deformation parameter $a$. It is given by
\begin{eqnarray}
S_\phi &=& \int
\d^4x\sqrt{-g}\,
\Big( g^{\mu\nu}D_\mu\phi^+D_\nu\phi -\mu^2\phi^+\phi \label{SExp}\\
&& + \frac{\theta^{\alpha\beta}}{2}g^{\mu\nu}\big( -\frac{1}{2}D_\mu\phi^+F_{\alpha\beta}
D_\nu\phi
+(D_\mu\phi^+)F_{\alpha\nu}D_\beta\phi + (D_\beta\phi^+)F_{\alpha\mu}D_\nu\phi\big) \Big)
.\nn 
\end{eqnarray}
The equation of motion for the field $\phi$ is obtained by varying this action with respect to $\phi$ and it is given by
\begin{equation}
\label{EoMPhi}
\begin{split} 
& g^{\mu \nu} \bigg( (\partial_{\mu} - iA_{\mu})D_{\nu} \phi - \Gamma^{\lambda}_{\mu
\nu}D_{\lambda} \phi \bigg) \\
& -\frac{1}{4} \theta^{\alpha \beta} g^{\mu \nu} \bigg(  (\partial_{\mu} - iA_{\mu})(
F_{\alpha \beta}D_{\nu} \phi)
- \Gamma^{\lambda}_{\mu \nu} F_{\alpha \beta}  D_{\lambda} \phi \\
& - 2  (\partial_{\mu} - iA_{\mu})( F_{\alpha \nu}D_{\beta} \phi)  + 2
\Gamma^{\lambda}_{\mu \nu} F_{\alpha \lambda}  D_{\beta} \phi  -2  (\partial_{\beta} -
iA_{\beta})( F_{\alpha \mu}D_{\nu} \phi) \bigg) = 0,  
\end{split}
\end{equation}
with $ \Gamma^{\lambda}_{\mu \nu}$ being the Christoffel symbols corresponding to the
metric (\ref{dsRN}). The RN background also fixes the $U(1)$ gauge field $A_\mu$ to 
\begin{equation}
A_0 = -\frac{qQ}{r}. \label{A0}
\end{equation}
This is the electromagnetic potential of the point-like charge $Q$ located at $r=0$. Consequently,
the only non-zero component of the field strength tensor $F_{\mu\nu}$ is the radial electric field
\begin{equation}
F_{r0} = \frac{qQ}{r^2}. \label{Fr0}
\end{equation}

Furthermore, since the only non-zero components of the NC deformation parameter
$\theta^{\alpha\beta}$ are
$\theta^{t\varphi}=
-\theta^{\varphi t}=a$, the equation of motion (\ref{EoMPhi}) simplifies to
\begin{eqnarray}
&&
\Big( \frac{1}{f}\partial^2_t -\Delta + (1-f)\partial_r^2 
+\frac{2MG}{r^2}\partial_r + 2iqQ\frac{1}{rf}\partial_t -\frac{q^2Q^2}{r^2f}\Big)\phi
\nonumber\\
&& +\frac{aqQ}{r^3}
\Big( (\frac{MG}{r}-\frac{GQ^2}{r^2})\partial_\varphi
+ rf\partial_r\partial_\varphi \Big) \phi =0, \label{EomPhiExp1}
\end{eqnarray}
where $\Delta$ is the usual Laplace operator. We also introduced
\begin{equation}
f=1-\frac{2MG}{r}+\frac{Q^2G}{r^2} = \frac{(r-r_+)(r-r_-)}{r^2}, \label{fRN} 
\end{equation}
with $r_{\pm}=MG \pm
\sqrt{M^2G^2-Q^2G}$. To simplify the notation, we will set $G=1$ in the following.

In order to solve this equation we assume an ansatz \cite{Ciric:2017rnf}
\begin{equation}
\phi_{lm}(t,r,\theta,\varphi) = R_{lm}(r)e^{-i\omega t}Y_l^m(\theta, \varphi) \label{AnsatzPhi} 
\end{equation}
with spherical harmonics $Y_l^m(\theta, \varphi)$. Inserting (\ref{AnsatzPhi}) into
(\ref{EomPhiExp1}) leads to an equation for the radial function $R_{lm}(r)$
\begin{eqnarray}
&& f R_{lm}'' + \frac{2}{r}\big( 1-\frac{M}{r}\big) R_{lm}' - \Big( \frac{l(l+1)}{r^2} 
- \frac{1}{f}(\omega - \frac{qQ}{r})^2   + \mu^2  \Big)R_{lm} \nn\\
&& -ima\frac{qQ}{r^3}\Big( (\frac{M}{r}-\frac{Q^2}{r^2})R_{lm} + rf R_{lm}' \Big) =0
.\label{EoMR} 
\end{eqnarray}
We note that the first line of this equation, which describes the system without deformation, corresponds to the equation for the radial function $R_{lm}$ analyzed in
\cite{Hod:2010hw, HodWKBRN}. The NC contribution in (\ref{EoMR}) vanishes for a neutral
scalar field, that is for $q=0$, while it does not depend on the scalar field mass $\mu$.

\section{QNM spectrum: WKB analysis}

Quasinormal modes are a particular solution of equation (\ref{EoMR}). They are specified by the
following boundary conditions: purely incoming at the horizon and purely outgoing in the infinity.
We mentioned in the Introduction that there are different ways to solve equation (\ref{EoMR}) and
find the corresponding QNM spectrum. To warm up, we start with the WKB approach and present an
analytic solution for the QNM frequencies, valid for a specific range of parameters. Later on, in
Section 4 we move to a more general numerical method, the continued fraction method.

\subsection{A modified tortoise coordinate}

The starting point of the WKB method is a Schr\"{o}dinger type equation
\begin{equation} \label{schrod}
\frac{\d^2 \psi}{\d {r_*}^2}  + V \psi =0.
\end{equation}
Therefore, we have to transform equation (\ref{EoMR}) into this form. To start with, we have to
define the tortoise coordinate $r_*$. However, the usual definition of $r_*$
\begin{equation}
\d r_*= \frac{\d r}{f} \label{TortoiseUsual}
\end{equation}
does not lead to an equation of the type (\ref{schrod}). The form of the equation is spoiled by the presence of terms that are linear in $\frac{\d \psi}{\d r_*}$. The origin of these terms is the presence of the deformation parameter $a$. Fortunately, it is possible to circumvent this problem by conveniently extending the definition of the tortoise coordinate.
Indeed, one can show that the following change of coordinates
\begin{equation} \label{modtortoise}
\d y = \frac{ \d r}{f \bigg( 1+ iam \frac{qQ}{r} \bigg)}
\end{equation}
brings the equation (\ref{EoMR}) into the form
\begin{equation} \label{schrod2}
\frac{\d^2 \psi}{\d {y}^2}  + V \psi =0,
\end{equation}
where $\psi=rR$ and
\begin{eqnarray} \label{veffektivno}
V &=& rf\Bigg[ -\frac{2}{r^3} \big(  \frac{M}{r} - \frac{Q^2}{r^2} \big) - \frac{1}{r} \Bigg( \frac{l(l+1)}{r^2} -
\frac{1}{f} {\bigg( \omega - \frac{qQ}{r}   \bigg)}^2  + \mu^2 \Bigg)   \\
&+&  iam \frac{qQ}{r^4} \big( 1 - \frac{7M}{r}    + 6\frac{Q^2}{r^2}  \big) - 2iam \frac{qQ}{r^2}\Bigg( \frac{l(l+1)}{r^2} 
-\frac{1}{f} {\bigg( \omega - \frac{qQ}{r}   \bigg)}^2  + \mu^2 \Bigg)  \Bigg]. \nonumber
\end{eqnarray}
Note that, like in all our calculations, this equation is valid up to first order in the deformation parameter $a$.

It is interesting to see what is the explicit form of the tortoise coordinate defined by (\ref{modtortoise}).
In order to see this, we note that up to first order in $a$ the relation (\ref{modtortoise})  can equivalently be written as
\begin{equation}  \label{tortint}
y= y^{(0)} + y^{(1)} = \int \frac{\d r}{f} -ia m qQ ~\int \frac{\d r}{r f},
\end{equation}
clearly separating a required transformation into two parts. The first part is the standard
Reissner–Nordstr\" om tortoise coordinate 
$y^{(0)} = r_*^{RN}$ while the second part represents the term coming exclusively from the
NC deformation.

Integrating (\ref{tortint}) we find
\begin{eqnarray} \label{modtortoise1}
y &=& y^{(0)} -ia m qQ  ~ \Bigg \{  \frac{r_+}{r_+ - r_-} \ln (r- r_+) - \frac{r_-}{r_+ - r_-} \ln (r- r_-) \Bigg \},  \\
&=& r + \frac{r_+}{r_+ - r_-} \Big(r_+ - iam qQ \Big) \ln (r- r_+) - \frac{r_-}{r_+ - r_-} \Big(r_- - iamqQ \Big) \ln (r- r_-)  ,\nonumber
\end{eqnarray}
where $y^{(0)}$ is the standard tortoise coordinate for the Reissner–Nordstr\" om metric given by
\begin{equation}  
y^{(0)} \equiv r_*^{RN} = r + \frac{r^2_+}{r_+ - r_-} \ln (r- r_+) - \frac{r^2_-}{r_+ - r_-} \ln (r- r_-). \label{StandardTortCoord}
\end{equation}
We repeat once again that the result (\ref{modtortoise1}) is valid up to first order in the parameter $a$.
The corrections in (\ref{modtortoise1}) are such that they do not change the position of the event  horizons, which coincides with the  analysis of the Hawking radiation in a semiclassical tunneling formalism \cite{wilczek,Ciric:2017rnf}.

The coordinate $y$ (\ref{modtortoise1}) has the standard properties of a tortoise coordinate. As
$r \rightarrow +\infty, ~ y \rightarrow +\infty $, and as 
$r \rightarrow r_+, ~ y \rightarrow -\infty$.
Moreover, a brief inspection of the  potential  (\ref{veffektivno})  shows that as $~r \rightarrow +\infty, ~$ the  potential $V$  tends to
$~V  \rightarrow   \omega^2 -  {\mu}^2.$ Similarly, as $~r \rightarrow r_+, ~$ the  potential $V$  approaches the value
$~V  \rightarrow    {\Big( \omega - \frac{qQ}{r_+} \Big)}^2 \Big( 1 - 2ia m \frac{qQ}{r_+} \Big)$. These two limiting values completely agree with the behavior of the effective potential
found in \cite{Konoplya:2013rxa} for the massive charged scalar field in the Kerr-Newman background (after putting the black hole angular momentum of the reference \cite{Konoplya:2013rxa} to zero and taking the commutative limit of our result).

\subsection{QNM spectrum}

The WKB method is based on the similarity between the equation governing the behavior of a black hole perturbation (scalar in our case) 
and the Schr\" odinger equation in the case of a potential barrier \cite{schutzwill,iyer,mashoon}. The condition for QNMs is obtained by matching two WKB solutions on each side of the potential barrier given by $-V$ to the solution inside the barrier, with the matching done simultaneously across both of the turning points. 
The matching procedure \cite{KonopReview} leads to the following condition for the QNM frequencies:
\begin{equation}  \label{wkbQNMcondition}
-\frac{V_0}{\sqrt{2V''_0}} -i \sum_{j=2}^6 \Lambda_j = i \big(n + \frac{1}{2} \big),
\end{equation}
where $n=0,1,2,3,...$. This condition clearly involves finding an extremal value of the  potential $V$ and the value of the curvature at the extremal point \cite{schutzwill, iyer}. In addition, the correction terms $\Lambda_j$  depend on the value of
the effective potential and its derivatives  (up to $j$-th order) in the maximum. The explicit form of the WKB
corrections $\Lambda_2$ and $\Lambda_3$ can be found in \cite{iyer}  and of $\Lambda_4, \Lambda_5,
\Lambda_6$ in \cite{konoplya}. Here, we will work in the 1st WKB order. Due to the additional terms
in (\ref{veffektivno}) induced by noncommutativity, the higher WKB corrections are very cumbersome
to calculate. However, these corrections are small and they will not influence the leading order
behavior. Therefore, we postpone the analysis of the higher WKB corrections for our future work.

In order to proceed we follow \cite{HodWKBRN} and introduce new dimensionless variables
\begin{equation} \label{dimensionlessvariables}
x = \frac{r - r_+}{r}, \qquad  \Omega = \frac{\omega r_+}{qQ} -1.
\end{equation}
In terms of these variables the effective potential can be written as
\begin{eqnarray} \label{veffektivno2}
V(x) &=& - V_{eff} =   {\bigg( \frac{qQ}{r_+}  \bigg) }^2 {(x + \Omega})^2 
- \frac{r_+ - r_-}{r_+} \Big(\mu^2 + \frac{H(r_+)}{r_+^2}\Big) x \nonumber \\
&& +  iam \frac{qQ}{r_+}  \frac{r_+ - r_-}{r_+} \Big(    \frac{1}{r_+^2} - \frac{7M}{r_+^3}
+ 6\frac{Q^2}{r_+^4} - 2\frac{l(l+1)}{r_+^2} - 2\mu^2  \Big) x \nn
\end{eqnarray}
\begin{eqnarray}
&& +4iam\Big(\frac{qQ}{r_+}\Big)^3\Omega x + 2iam \Big( \frac{qQ}{r_+} 
\Big)^3 \Omega^2 (1-x)  + O(x^2, ax^2),\nonumber  
\end{eqnarray}
where $H(r_+) = l(l+1) + \frac{2M}{r_+} - \frac{2Q^2}{r_+^2} = l(l+1) + \frac{r_+ - r_-}{r_+}$.  
It is obvious from these definitions that  $H(r_+) < l(l+1) + 1$. Note that we discarded all terms quadratic in $x$ other than the first parabolic term  $\sim {(x+\Omega)}^2$ as non physical.  Indeed,  if we keep all quadratic terms, the resulting effective potential does not describe a realistic problem at hand. In particular, within our approximation the parabolic term  $\sim {(x+\Omega)}^2$ has a peak immediately next to the horizon. Retaining other terms quadratic in $x$ would only move the peak of the potential far away from the event horizon and this is physically undesirable. That is why, of all terms quadratic in $x$, we keep only the first parabolic term $\sim {(x+\Omega)}^2$ in the effective potential. Moreover, retaining  terms of the order $O(ax^2)$ in the potential function would only produce corrections to the position of its  extremal point that are of the second order within our approximation and thus can be neglected.

As far as the effective potential of the Reissner–Nordstr\" om
black hole  due to  charged scalar perturbations is concerned, the analysis carried out in reference \cite{HodWKBRN}
has been focused on the regime $qQ \gg l+1$ with a purpose  of being able  to deal  with the condition (\ref{wkbQNMcondition}) analytically 
in that particular case. For the same reason we also restrict our analysis to this regime.
Putting this together with the just established relation  $H(r_+) < l(l+1) + 1$,  
it is clear that this condition translates into  $\frac{H(r_+)}{q^2 Q^2}\ll 1 $. 
In addition, we also assume that $\frac{\mu^2 r_+^2}{q^2 Q^2} \ll 1$. 

It is reasonable to expect the QNM frequencies to be centered around the classical result
$\omega = \frac{qQ}{r_+}$. Therefore, it is  natural to assume that the quantity $\Omega$ is also very small. As in the reference \cite{HodWKBRN},
we assume here that it is of the same order as $\frac{H(r_+)}{q^2 Q^2}$ and $\frac{\mu^2 r_+^2}{q^2 Q^2}$. Furthermore, our analysis includes a small NC parameter $a$ and we assume that it is also of the same order as the former small quantities. To be more precise, we have to assume that $la\frac{qQ}{r_+} \ll 1$, where $l$ is the orbital momentum of the perturbation. To summarize, our approximation can be written as: $\frac{H(r_+)}{q^2 Q^2} \sim \frac{\mu^2 r_+^2}{q^2 Q^2} \sim \Omega \sim la\frac{qQ}{r_+} \ll 1$.

The position of the extremum of the effective potential is determined by the condition $\frac{\d V}{\d x}\Big|_{x_0} =0$, leading to
\begin{align}
x_0 + \Omega = \hspace{2mm}& \frac{1}{2 q^2 Q^2} \frac{r_+ - r_-}{r_+} \Big[ H(r_+) + \mu^2 r_+^2    \label{xNula} \\
& - iam \frac{qQ}{r_+} \Big( 1- \frac{7M}{r_+} + \frac{6Q^2}{r_+^2} - 2l(l+1) - 2\mu^2  r_+^2 
\Big) \Big]\nn\\
&  - 2iam \frac{q^3 Q^3}{r_+} (2\Omega - \Omega^2) .\nn
\end{align}
It is easily seen that within our approximation, the value of $x_0= \frac{r_0-r_+}{r_0}$ is very small, ensuing that the peak of the potential is close to the event horizon, as it should be.
The value of $r_0$ can be obtained from the relation $x_0 = 1-\frac{r_+}{r_0}$, leading to
\begin{align}
\frac{1}{r_0} = \hspace{2mm}&  \frac{\Omega + 1}{r_+} -   \frac{1}{2 q^2 Q^2}   \frac{r_+ - r_-}{r^2_+} \Big[ H(r_+) + \mu^2 r_+^2  \nn\\
&  - iam \frac{qQ}{r_+} \Big( 1- \frac{7M}{r_+} + \frac{6Q^2}{r_+^2} - 2l(l+1) - 2\mu^2  r_+^2  \Big) \Big]  
+ 2iam \frac{q^3 Q^3}{r^2_+} (2\Omega - \Omega^2). \nonumber
\end{align}
Up to first order in $\frac{H(r_+)}{q^2 Q^2}, \frac{\mu^2 r^2_+}{q^2 Q^2}, \Omega$ and $la\frac{qQ}{r_+}$ this equation can be inverted to give
\begin{align}
r_0 = r_+ \Bigg[ 1- \Omega + (r_+ - r_-)  \Big[ & \frac{H(r_+)}{2q^2 Q^2 r_+} +  \frac{\mu^2 r_+}{2q^2 Q^2} \label{rnula}\\
&- iam  \frac{1}{2qQ r^2_+} \Big( 1- \frac{7M}{r_+} + \frac{6Q^2}{r_+^2} - 2l(l+1) - 2\mu^2  r_+^2  \Big) \Big]  \Bigg].\nn
\end{align}
In the same way one finds the remaining quantities that will be required in the analysis, see the relation (\ref{wkbQNMcondition1}):
\begin{equation} \label{dxdrrnula}
\begin{split}
{\Big( \frac{\d x}{\d r} \Big)}_{r= r_0} =  \frac{1}{r_+}  \Bigg[& 1+ 2 \Omega - \frac{1}{q^2 Q^2} \frac{r_+ - r_-}{r_+} 
\Big[ H(r_+)  +  \mu^2 r^2_+  \nn\\
&- iam \frac{qQ}{r_+}\Big( 1- \frac{7M}{r_+} + \frac{6Q^2}{r_+^2} - 2l(l+1) - 2\mu^2  r_+^2  \Big) \Big]  \Bigg]
\end{split}
\end{equation}
and
\begin{eqnarray} 
f(r_0)  &=& \frac{(r_0 - r_+)(r_0 - r_-)}{r^2_0} \nn\\
&=& \frac{r_+ - r_-}{r^2_+} \Bigg[r_+ (r_+ - r_-)\Big[  \frac{H(r_+)}{2q^2 Q^2 r_+} +  \frac{\mu^2 r_+}{2q^2 Q^2}  \label{fodrnula}  \nonumber \\
&&- iam  \frac{1}{2qQ r^2_+} \Big( 1- \frac{7M}{r_+} + \frac{6Q^2}{r_+^2} - 2l(l+1) - 2\mu^2  r_+^2  \Big)  \Big] - \Omega r_+   \Bigg] .
\end{eqnarray}
The extremal value $V_0$ of the negative effective potential $V(x) = - V_{eff}$ is
\begin{align}
V_0 = \hspace{2mm}& -\frac{1}{4q^2 Q^2 r^2_+ }
 {\bigg( \frac{r_+ - r_-}{r_+} \bigg)}^2 {\Big( H(r_+) + \mu^2 r_+^2 \Big) }^2  \nn\\
& + \mu^2 \Omega  \frac{r_+ - r_-}{r_+} +     \frac{H(r_+) (r_+ - r_-)}{r^3_+}  \Omega .\label{wkbQNMcondition2}
\end{align}	
Once again, note that all these quantities have been written only up to first order in $\frac{H}{q^2 Q^2}$, $\frac{\mu^2}{q^2 Q^2}$, $\Omega$ and $la\frac{qQ}{r_+}$.

Since at the extremal point $x_0$ the first derivative $\frac{\d V}{\d x}$ vanishes, the QNM condition (\ref{wkbQNMcondition})
implies
\begin{equation}
\label{wkbQNMcondition1}
-V_0 = i \big( n+\frac{1}{2} \big) \sqrt{2  {\bigg( \frac{\d^2 V}{\d x^2} \bigg)}_{x=x_0}  {\bigg(\frac{\d x}{\d r}\bigg) }^2_{r=r_0}{\bigg( \frac{\d r}{\d y} \bigg)}^2_{y=y_0}}.
\end{equation}
Knowing that $\frac{\d r}{\d y} = f(r)  \big(1+iam \frac{qQ}{r} \big)$ and inserting the above expressions for $\Big(\frac{\d x}{\d r}\Big)_{r=r_0}$, $V_0 $ and $f(r_0)$ into  (\ref{wkbQNMcondition1}) leads to a coupled system of equations
\begin{eqnarray} 
&& \frac{(r_+ - r_-)}{4q^2 Q^2 r_+} \Big( H(r_+) + \mu^2 r_+^2 \Big)^2 - \Omega_R \Big[  H(r_+) + \mu^2 r_+^2  \Big]   \label{wkbQNMcondition3a}  \\
&& =  2qQ\big( n + \frac{1}{2} \big) \Omega_I + am \big( n + \frac{1}{2} \big) \frac{r_+ - r_-}{r^2_+}
\Big( 1- \frac{7M}{r_+} + \frac{6Q^2}{r_+^2} - 2l(l+1) - 2\mu^2  r_+^2  \Big),  \nn \\
&& - \Omega_I \Big[  H(r_+) + \mu^2 r_+^2  \Big] \nn\\
&&=  \big( n + \frac{1}{2} \big) \frac{r_+ - r_-}{qQr_+} \Big[  H(r_+) + \mu^2 r_+^2  \Big] - 2qQ  \big( n + \frac{1}{2} \big) \Omega_R,  \label{wkbQNMcondition3b}
\end{eqnarray}
where $\Omega$ has been split in its real and imaginary part, $\Omega = \Omega_R + i\Omega_I$. The solution of this system of equations is given by
\begin{eqnarray}
\Omega_R &=& \frac{(r_+ -r_-) (H(r_+) + \mu^2 r^2_+)}{r_+ \Big( (H(r_+) + \mu^2 r^2_+)^2 + 4q^2Q^2(n+\frac{1}{2})^2\Big)} \Bigg[ \frac{(H(r_+) + \mu^2 r^2_+)^2}{4q^2Q^2} \label{OmegaR}\\ 
&& + 2 (n+\frac{1}{2})^2 -\frac{am}{r_+}(n+\frac{1}{2}) \Big( 1- \frac{7M}{r_+} + \frac{6Q^2}{r_+^2} - 2l(l+1) - 2\mu^2  r_+^2 \Big) \Bigg], \nn\\
\Omega_I &=& \frac{(r_+ -r_-)}{2qQr_+}\Bigg[-\frac{(H(r_+) + \mu^2 r^2_+)^2}{4q^2Q^2} \label{OmegaI}\\
&& -\frac{am}{r_+}(n+\frac{1}{2}) \Big( 1- \frac{7M}{r_+} + \frac{6Q^2}{r_+^2} - 2l(l+1) - 2\mu^2  r_+^2 \Big)\Bigg] .\nn
\end{eqnarray}
Using (\ref{dimensionlessvariables}) one can express the real and the  imaginary part of the QNM
frequency $\omega$. For completeness, in Figure 1 we plot the dependence of Re\,$\omega$ and
Im\,$\omega$ on $qQ$ in the case of $\mu=0.05$, $l=100$  and $a=0.000001$. It is obvious from
equation (\ref{OmegaI}) that there will be a splitting of frequencies for different values of the
projection of angular momentum $m$. This result is in agreement with our previous findings in
\cite{Ciric:2017rnf}. The frequency splittings $\omega^\pm = \omega (m=\pm 100) - \omega (m=0)$ are
plotted in Figure 2.

The straightforward comparison of the results obtained by the  WKB method with our previous results
in \cite{Ciric:2017rnf} obtained analytically in the near-extremal limit in not possible. A well
known feature of the WKB method is that it gives better results for higher values of the angular
momentum $l$. Therefore we plotted our results for the case $l=100$. On the other hand, in
\cite{Ciric:2017rnf} we analyzed only $l=1$ and $l=2$ cases. However, the qualitative comparison
shows the frequency splitting due to noncommutativity in both cases and confirms our conclusions
about the effects of noncommutative deformation on the QNM spectrum. Notice that, unlike in
\cite{Ciric:2017rnf}, the splitting in Im\,$\omega$ is already visible in Figure 1. Namely, we
plotted three cases $m=-100$, $m=0$ and $m=100$. Since the NC corrections are proportional to $am$,
it is obvious that the effect of noncommutativity will be larger for bigger $m$. The relative
splitting can be estimated from the Figures 1-4 as $\delta_{\mbox {\tiny Re}} \sim \frac{{\mbox
Re}\,\omega^+}{{\mbox Re}\,\omega} \sim 10^{-4}$. In the similar way, for the imaginary part we
estimate $\delta_{\mbox {\tiny Im}} \sim \frac{{\mbox Im}\,\omega^+}{{\mbox Im}\,\omega} \sim
10^{-2}$. The splitting is obviously bigger then the splitting found in \cite{Ciric:2017rnf} and
this enhancement is due to the high angular momentum $l=100$.

\begin{center}
\begin{tabular}{lll}
\includegraphics[scale=0.33]{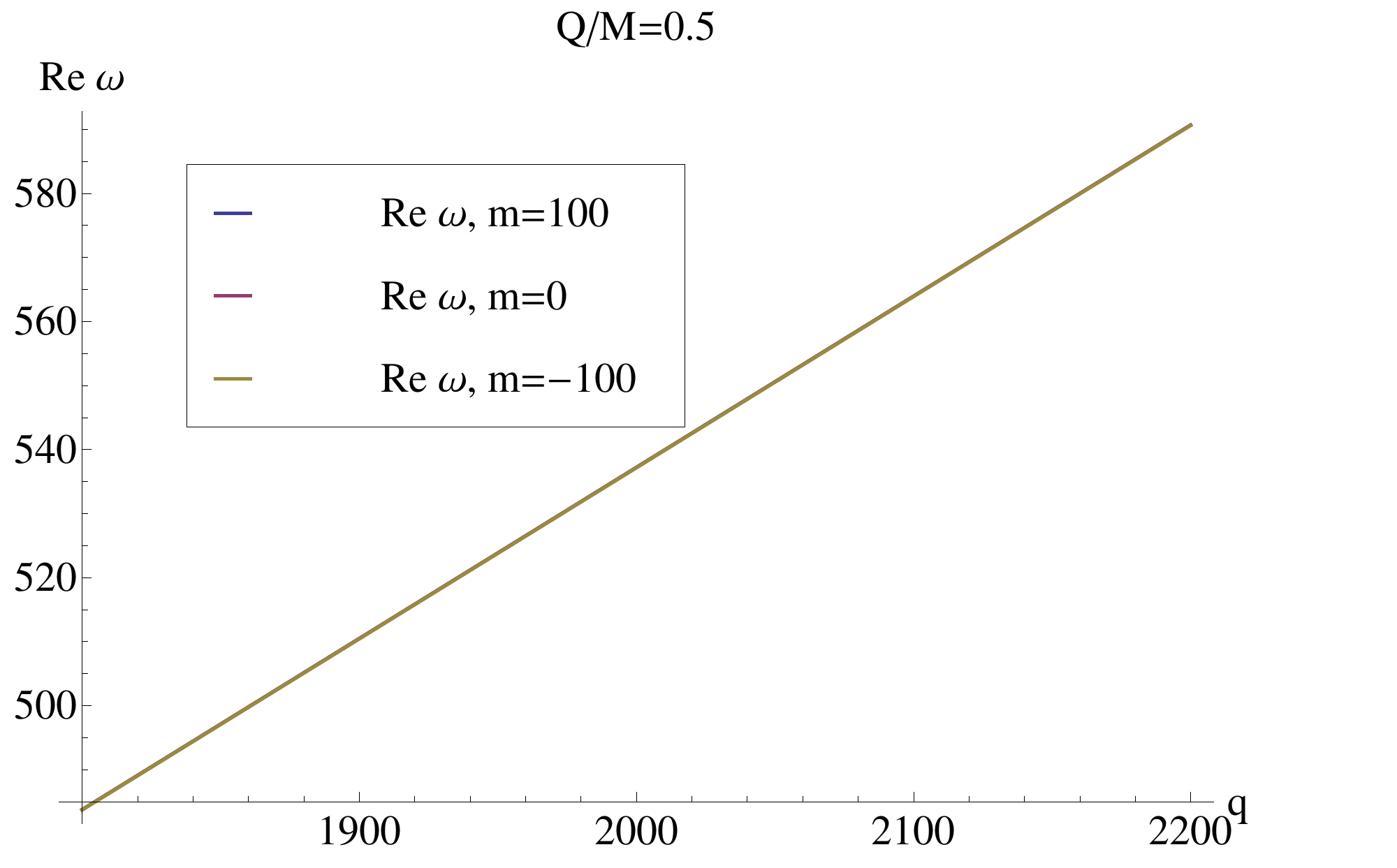}& & \hspace{-3mm}\includegraphics[scale=0.33]{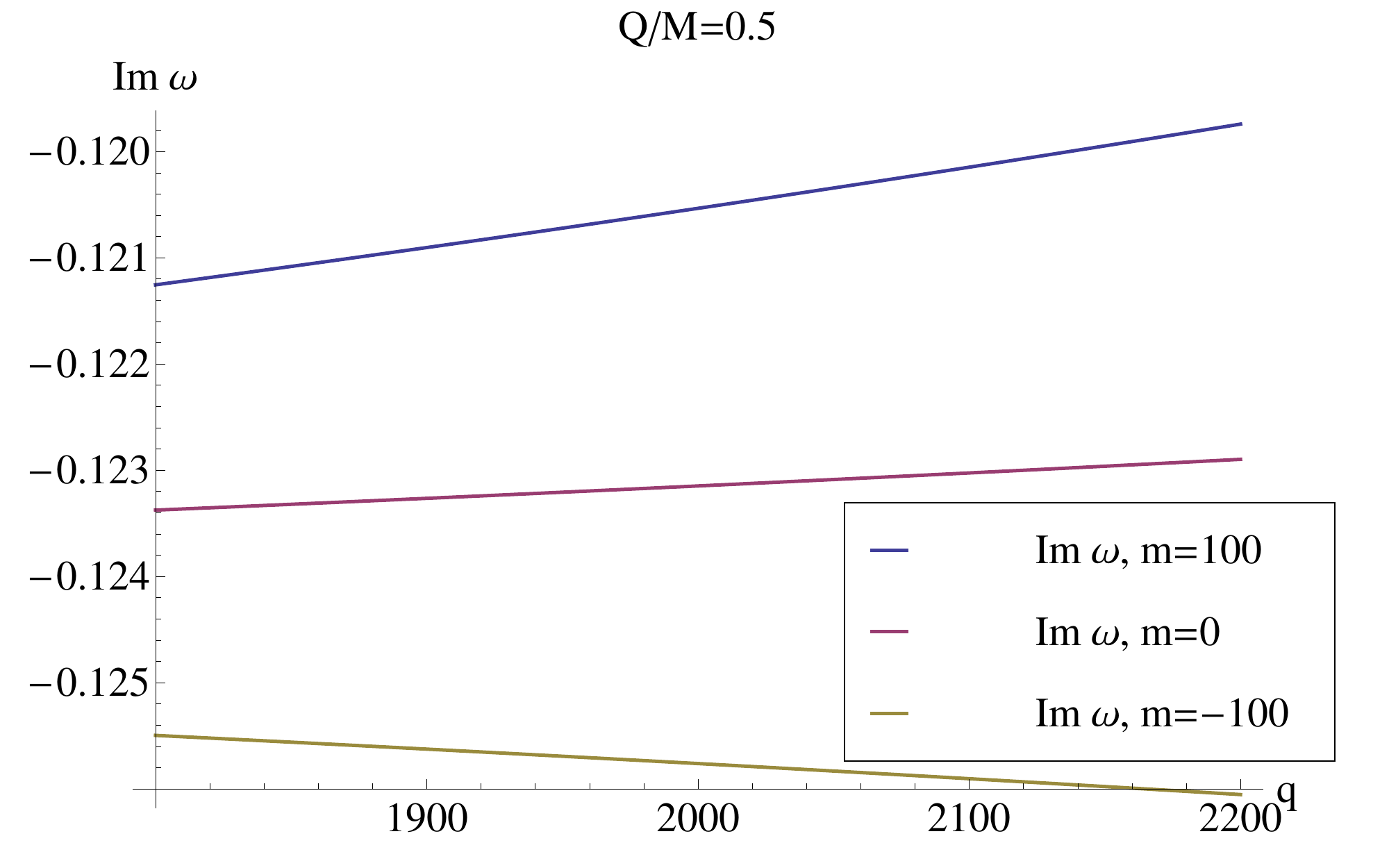}\\
\multicolumn{3}{l}{{\scriptsize {\bf Figure 1:} Dependence of Re\,$\omega$ (left) and Im\,$\omega$ (right) on the charge $qQ$ of the scalar field with the mass }}\\ 
\multicolumn{3}{l}{{\scriptsize $\mu=0.05$ and the orbital momentum $l=100$.}}
\end{tabular} 
\end{center}
\vspace{3mm}
\begin{center}
\begin{tabular}{lll}
\includegraphics[scale=0.33]{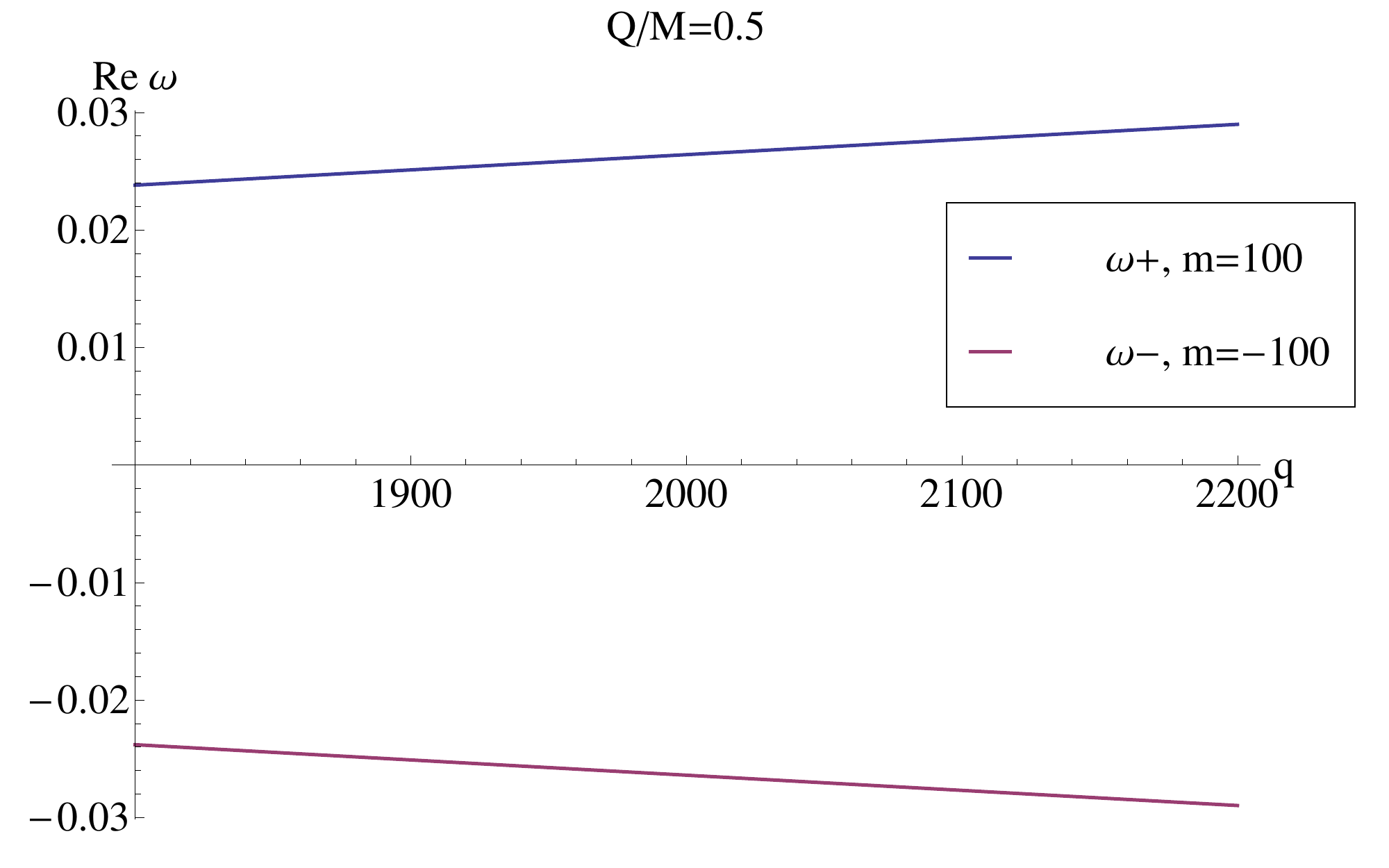} & & \hspace{-3mm}\includegraphics[scale=0.33]{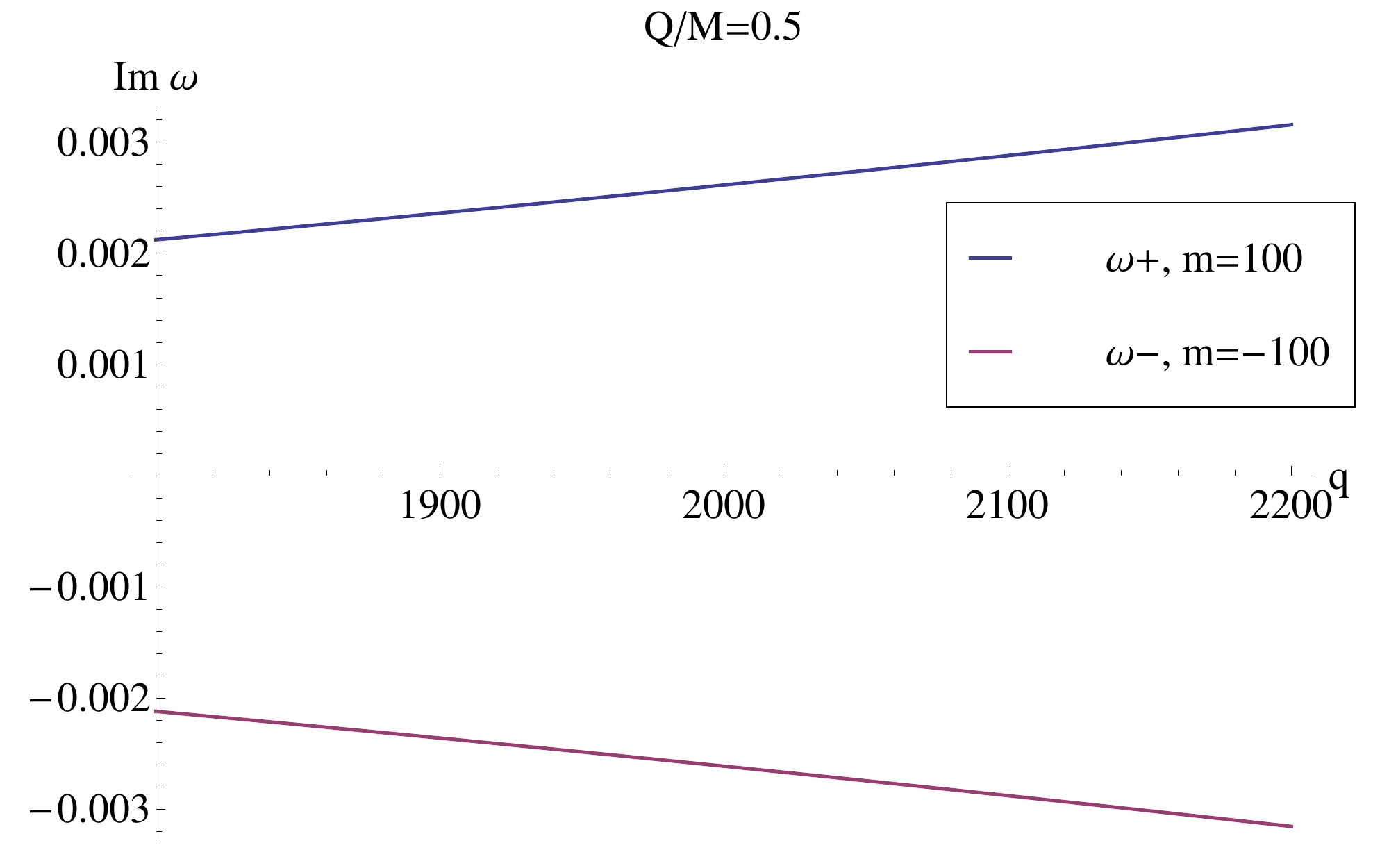}\\
\multicolumn{3}{l}{\scriptsize {\bf Figure 2:} Dependence of Re\,$\omega^\pm$ (left) and
Im\,$\omega^\pm$ (right) on the charge $qQ$ of the scalar field with }\\
\multicolumn{3}{l}{\scriptsize the mass $\mu=0.05$ and $m=\pm 100$.}
\end{tabular} 
\end{center}

Finally, to compare our results with the commutative results obtained in \cite{HodWKBRN}, we have to
make one additional approximation. Namely, we assume that $ \frac{ q^2 Q^2  }{(H(r_+) + \mu^2
r^2_+)^2} \ll 1$, and at the same time being of the same order of magnitude as $\frac{H(r_+)}{q^2
Q^2}$, $\frac{\mu^2 r^2_+}{q^2 Q^2}$, $\Omega$ and $al\frac{qQ}{r_+}$. This final approximation
amounts to requiring $q^2 Q^2 \ll l^4$. Taken together, our approximations select the parameter
range to be $l \ll qQ \ll l^2$.

Solving the above system of equations for $\Omega_R$ and $\Omega_I$ then  leads respectively to
\begin{eqnarray} 
\Omega_R &=&
\frac{r_+ - r_-}{4q^2Q^2 r_+}(H(r_+) + \mu^2 r^2_+) \Bigg[ 1 + \frac{8q^2Q^2}{(H(r_+) + \mu^2 r^2_+)^2}(n+\frac{1}{2})^2\nn\\
&& -\frac{4q^2Q^2}{r_+(H(r_+) + \mu^2 r^2_+)^2} am(n+\frac{1}{2})\Big( 1- \frac{7M}{r_+} + \frac{6Q^2}{r_+^2} - 2l(l+1) - 2\mu^2  r_+^2  \Big)\Bigg]\nn\\
&\approx& \frac{r_+ - r_-}{4q^2Q^2 r_+}(H(r_+) + \mu^2 r^2_+) \Big[ 1 + \frac{8q^2Q^2}{(H(r_+) + \mu^2 r^2_+)^2}(n+\frac{1}{2})^2 \Big] \label{wkbQNMcondition4}
\end{eqnarray}
and
\begin{eqnarray} 
\Omega_I &=& -\frac{r_+ - r_-}{qQ r_+} \big( n + \frac{1}{2} \big)
+ \frac{1}{(H(r_+) + \mu^2 r^2_+)^2}   \frac{r_+ - r_-}{ r_+}    \big( n + \frac{1}{2} \big)\times   \nn\\
&&\Bigg[ \frac{1}{2 q Q }\Big( H(r_+) + \mu^2 r^2_+  \Big)^2 + 4qQ {\big( n + \frac{1}{2} \big)}^2  \label{wkbQNMcondition5}\\
&& - 2am \frac{qQ}{r_+}  \big( n + \frac{1}{2} \big)
\Big( 1- \frac{7M}{r_+} + \frac{6Q^2}{r_+^2} - 2l(l+1) - 2\mu^2  r_+^2  \Big) \Bigg].    \nonumber  \\
&\approx& -\frac{r_+ - r_-}{qQ r_+} \big( n + \frac{1}{2} \big)
+ \frac{1}{(H(r_+) + \mu^2 r^2_+)^2}   \frac{r_+ - r_-}{ r_+}    \big( n + \frac{1}{2} \big)\times   \nn\\
&&\Bigg[ \frac{1}{2 q Q }\Big( H(r_+) + \mu^2 r^2_+  \Big)^2 + 4qQ {\big( n + \frac{1}{2} \big)}^2 \Bigg].\nn
\end{eqnarray}
Within this approximation, there is no contribution of noncommutativity in both $\Omega_R$ and
$\Omega_I$. The terms in the second line in (\ref{wkbQNMcondition4}) and
(\ref{wkbQNMcondition5}) are second order "small" being a product\footnote{Note that $qQ$ is always
bigger then $1$, since $l\ll qQ \ll l^2$. According to our approximation $al\frac{qQ}{r_+}\ll 1$,
therefore $al\frac{1}{r_+}\ll 1$ is also valid. In the same way $ \frac{ q^2 Q^2  }{(H(r_+) + \mu^2
r^2_+)^2} \ll 1$ implies $\frac{1}{(H(r_+) + \mu^2 r^2_+)^2} \ll 1$.} of $\frac{q^2Q^2}{(H(r_+) +
\mu^2 r^2_+)^2}$ and $am\frac{1}{r_+}$ in (\ref{wkbQNMcondition4}) and $\frac{1}{(H(r_+) + \mu^2
r^2_+)^2}$ and $am\frac{qQ}{r_+}$ in (\ref{wkbQNMcondition5}). To compare the obtained results with
the results in \cite{HodWKBRN}, we have to set the mass of the field $\mu=0$ and the
noncommutativity parameter $a=0$. After this, we indeed find the agreement with (19) and (20) in
\cite{HodWKBRN}. However, the results in \cite{HodWKBRN} are zeroth order in small variables $\frac{
q^2 Q^2  }{H^2}$, $\frac{H}{q^2 Q^2}$, $\Omega$, while our results also contain the first order
corrections in these variables.

\section{Continued fraction method}

So far we used two different methods to calculate the QNM spectrum of NC scalar perturbation of the RN geometry. Both methods have their limitations. The WKB method works well for high values of the orbital momentum $l$. The analytic treatment in \cite{Ciric:2017rnf} was possible only in the near-extremal limit of the RN geometry.

In this section we implement the continued fraction method \cite{leaver, nollert}
to determine the QNM spectrum of a massive charged scalar field around the RN black
hole in the presence of the noncommutative deformation of space-time. This method is less restrictive then the previous two and we expect to find results for a wider range of parameters. We mention that the analysis of the undeformed (commutative) (un)charged scalar and Dirac QNM spectrum in the RN background by the continued fraction method can be found in \cite{Konoplya:2013rxa, QNMRNBrazilci, Chowdhury:2018izv}. To our knowledge, this is the first time that the continued fraction method is applied also to the NC deformations of the commutative case.

We start by looking for the asymptotic form of the QNM spectrum in the spatial infinity $r\to \infty$ and near the horizon $r\to r_+$. The asymptotic form
can be obtained by analytically solving equation (\ref{EoMR}) in the
asymptotic limits of the spatial infinity  $r \rightarrow \infty$  and the event horizon  $r \rightarrow r_+$ and imposing the QNM boundary condition of purely incoming waves on the horizon and purely outgoing waves in the infinity. 

In the $r \rightarrow \infty$  limit equation (\ref{EoMR}) reduces to
\begin{equation} \label{eqasympfaraway}
\frac{ \d^2 \psi}{ \d y^2} + \Big[  \omega^2 - \mu^2 - 2 \frac{ \omega qQ - \mu^2 M }{r} + 2iam\frac{qQ}{r} \big(\omega^2 - \mu^2 \big) \Big] \psi =0. 
\end{equation}
The solution to the equation (\ref{eqasympfaraway}) is given by
\begin{equation} \label{asympfaraway}
R = \frac{\psi}{r}  \sim e^{\pm i \Omega y} y^{-1-i \frac{\omega qQ - \mu^2  M}{\Omega} - amqQ \Omega}. 
\end{equation}
Note that, in the limit $r\to \infty$, the parameter $\Omega$ is fixed by the leading order solution
to $\Omega^2 = \omega^2 - \mu^2$. Analogously, in the near horizon limit $r \rightarrow r_+,$ the
associated equation of motion (\ref{schrod}) reduces to
\begin{equation} \label{eqasymphorizon}
\frac{ \d^2 \psi}{ \d y^2} + \Big[  \omega -  \frac{qQ }{r_+} \Big]^2  \Big[ 1+ 2iam\frac{qQ}{r_+}  \Big] \psi =0, 
\end{equation}
with the solution
\begin{equation} \label{asymphorizon}
\psi   \sim e^{\pm i \Big( \omega  - \frac{ qQ}{r_+} \Big)  \Big( 1 + iam  \frac{ qQ}{r_+}  \Big)y }. 
\end{equation}
Solutions (\ref{asympfaraway}) and (\ref{asymphorizon}) are perturbative in the NC parameter $a$ and are valid up to first order in $a$. The QNM boundary conditions, purely outgoing in the infinity and purely incoming at the horizon, select signs in (\ref{asympfaraway}) and (\ref{asymphorizon}). Finally, the asymptotic form of the quasinormal modes is given by
\begin{gather}    
R(r) \rightarrow 
\begin{cases}         
Z_{out} e^{i \Omega y} y^{-1-i \frac{\omega qQ - \mu^2  M}{\Omega} - amqQ \Omega}, & \text{for } r \rightarrow \infty,\>\> (y\rightarrow \infty) \\  
  \\
Z_{in}  e^{-i \Big( \omega  - \frac{ qQ}{r_+} \Big)  \Big( 1 + iam  \frac{ qQ}{r_+}  \Big)y },   & \text{for }r \rightarrow  r_+, \>\> (y\rightarrow -\infty)                
\end{cases} . \label{ncboundaryconditions}   
\end{gather}
Here $Z_{out}$ and  $Z_{in}$ are the amplitudes of the outgoing and ingoing waves, respectively, and they do not depend on $r$ (or $y$). In the special case of a massless scalar field  $\mu =0$ and a vanishing space-time deformation $a=0$, these asymptotic solutions
reduce to the asymptotic solutions of \cite{QNMRNBrazilci}. 

Equation (\ref{EoMR}) has an irregular singularity at $r=+\infty$ and three regular singularities at
$r=0$, $r=r_-$ and $r=r_+$. To implement Leaver's method we expand the solution in terms of powers series
around  $r = r_+$. Then the radial part of the scalar field looks as
\begin{equation}  \label{generalpowersolution}
R(r) = e^{i \Omega r}  {(r-r_-)}^{\epsilon} \sum_{n=0}^{\infty} a_n {\Big( \frac{r-r_+}{r-r_-} \Big)}^{n + \delta}.
\end{equation}
The parameters $\delta$ and $\epsilon$ are determined by demanding that the solution (\ref{generalpowersolution}) satisfies the  boundary conditions (\ref{ncboundaryconditions}) at the horizon and in the infinity. From the general form (\ref{generalpowersolution}) of the solution, it is clear that as $r \rightarrow \infty$, the dominant behavior is determined by the term
$ r^{\epsilon} e^{i \Omega r} $. Likewise, as   $r \rightarrow r_+$, the dominant behavior of (\ref{generalpowersolution})  is given by the term $ {( r - r_+ )}^{\delta} $. On the other side,  we insert the expression  (\ref{modtortoise1})
for the tortoise coordinate    into (\ref{asympfaraway}) and (\ref{asymphorizon})
and then compare   the resulting expressions  with the formerly deduced asymptotic of (\ref{generalpowersolution}).  From this comparison the parameters $\epsilon$ and $\delta$
follow immediately and they are given by
\begin{equation}  \label{epsilondelta}
\delta = -i \frac{r_+^2}{r_+ - r_-} \Big( \omega - \frac{qQ}{r_+} \Big), \qquad
\epsilon = -1 - i qQ \frac{\omega}{\Omega} + i \frac{r_+ + r_-}{2\Omega} \Big(  \Omega^2 + \omega^2  \Big).
\end{equation}
It is worth noting that these parameters are the same as the corresponding ones in the reference \cite{Chowdhury:2018izv} and  thus are not affected by the noncommutative space-time deformation.

\subsection{Recurrence relations}

Now we insert the power series solution (\ref{generalpowersolution}) together with (\ref{epsilondelta}) into equation (\ref{EoMR}). In this way we obtain the recurrence relations for the coefficients $a_n$. The calculation leading to the recurrence relations is long and it is presented in the Appendix A. The result is the 6-term recurrence relation
\begin{eqnarray}  \label{6contfr}
A_n a_{n+1} + B_n a_n +C_n a_{n-1} + D_n a_{n-2} + E_n a_{n-3} + F_n a_{n-4 }  &=& 0,  n\geqslant 4\nonumber \\
A_3 a_{4} + B_3 a_3 +C_3 a_{2} + D_3 a_{1} + E_3 a_{0}   &=& 0, n=3\nonumber \\
A_2 a_{3} + B_2 a_2 +C_2 a_{1} + D_2 a_{0}   &=& 0, n=2\nonumber \\
A_1 a_{2} + B_1 a_1 +C_1 a_{0}    &=& 0, n=1\nonumber \\
A_0 a_{1} + B_0 a_0   &=& 0, n=0. 
\end{eqnarray}
The coefficients $A_n, B_n, C_n, D_n, E_n $ and $ F_n$ are given by
\begin{align}   
& A_n   =   r_+^3 \alpha_{n},  \label{contfr1} \\
&\nn\\
&  B_n  =  r_+^3 \beta_n - 3 r_+^2 r_- \alpha_{n-1}   \nonumber \\
&  \qquad -iamqQ (r_+ - r_-) r_+ (n + \delta) - \frac{1}{2} iamqQ (r_+ + r_-) r_+ +  iam qQ r_+ r_-
, \nonumber 
\end{align}
\begin{align}
& C_n  =  r_+^3 \gamma_n  + 3r_+ r_-^2 \alpha_{n-2}  -3r_+^2 r_- \beta_{n-1}\nn\\
& \qquad +iamqQ (r_+ - r_-)(2 r_+ +r_-) (n + \delta -1)  -iamqQ (r_+ - r_-)r_+ \epsilon   \nonumber \\ 
& \qquad +  \frac{1}{2}  iamqQ (r_+ + r_-)(2r_+ + r_-) - 3iamqQ r_+ r_- + amqQ \Omega {(r_+ -
r_-)}^2  r_+  ,  \nonumber \\
&\nn\\
& D_n  = - r_-^3 \alpha_{n-3}  + 3r_+ r_-^2 \beta_{n-2} -3 r_+^2 r_- \gamma_{n-1} +iamqQ (r_+^2 - r_-^2)\epsilon
+ 3iam qQ r_+ r_- \nonumber \\
&  \qquad -amqQ \Omega {(r_+ - r_-)}^2 r_- -iamqQ (r_+ - r_-) ( r_+ +2r_-) (n + \delta -2) \nn\\
&  \qquad -\frac{1}{2}  iamqQ (r_+ + r_-)(r_+ +2 r_-) ,  \nonumber \\
&\nn\\
&  E_n  =   3r_+ r_-^2 \gamma_{n-2} - r_-^3 \beta_{n-3} + iamqQ (r_+ - r_-) r_-  (n + \delta -3) \nonumber \\
&  \qquad     -iamqQ (r_+ - r_-)r_- \epsilon +  \frac{1}{2}  iamqQ (r_+ + r_-)r_- + iam qQ r_+ r_- ,
 \nonumber \\
&\nn\\
&  F_n = -r_-^3 \gamma_{n-3}.\nn
\end{align}
The coefficients $\alpha_n, \beta_n, \gamma_n$ are
\begin{align}
& \alpha_n   =  (n+1) \Big[ n + 1 -2i \frac{r_+}{r_+ - r_-} (\omega r_+  - qQ) \Big], \label{contfrsimple} \\
&  \beta_n  =   \epsilon + (n + \delta)(2 \epsilon - 2n -2\delta) + 2i\Omega (n + \delta) (r_+ - r_-) - l(l+1)  -\mu^2 r_-^2 \nonumber \\
&  \qquad     + \frac{2\omega r_-^2 }{r_+ - r_-} (\omega r_+ - qQ) - \frac{2 r_-^2 }{{(r_+ - r_-)}^2} {(\omega r_+ - qQ)}^2 + 4\omega r_- (\omega r_+ - qQ)    \nn \\
&  \qquad  -  \frac{2 r_- }{r_+ - r_-} {(\omega r_+ - qQ)}^2  +  (r_+ - r_-)  \Big[  i \Omega  + 2\omega (\omega r_+ - qQ)
   -  \mu^2   (r_+ + r_-)  \Big]  , \nonumber \\ 
&  \gamma_n =  \epsilon^2 + (n + \delta -1)(n+ \delta -1 -2\epsilon) + {\Big(  \omega r_- - \frac{r_-}{r_+ - r_-} (\omega r_+ - qQ)  \Big)}^2.     \nn
\end{align}
Let us clarify these relations. The first relation in (\ref{6contfr}), for $n\geqslant 4$, is
a general $6$-term recurrence relation. The remaining four relations are the indicial equations
relating the lowest order coefficients $a_n$ in the general expansion (\ref{generalpowersolution}).
They may be thought as boundary conditions for the  first relation in (\ref{6contfr}). The presence of NC deformation, through the terms linear in the NC parameter $a$, induces the 6-term recurrence relation. In the commutative case and for the non-extremal RN background the 3-term recurrence relation is obtained \cite{QNMRNBrazilci, Chowdhury:2018izv}. 

To compare the commutative limit of our result (\ref{6contfr}) with the results in the commutative
case, we have to  go back to the equation (\ref{appendix3}). There we impose the
commutative limit $a\to 0$ and divide by $(r_{+}-zr_{-})^3$ the remaining terms. The obtained
equation then results in the 3-term recurrence relation of \cite{QNMRNBrazilci, Chowdhury:2018izv},
\begin{eqnarray}    \label{3contfr}
\alpha_n a_{n+1} + \beta_n a_n +\gamma_n a_{n-1} = 0, \nonumber \\
\alpha_0 a_{1} + \beta_0 a_0   = 0,  
\end{eqnarray}
with the coefficients $\alpha_n$, $\beta_n$ and $\gamma_n$ given in (\ref{contfrsimple}). Note that the noncommutativity does not influence the parameters $\alpha_n$, $\beta_n$ and $\gamma_n$.

Having the recurrence relations that involve more than $3$ expansion coefficients $a_n$, as we do have here, we cannot straightforwardly apply the usual method for solving the recurrence relations \cite{gautschi}. Instead, we should first use the Gauss elimination method to gradually reduce the initial recurrence relation
from the $6$-term recurrence relation to a $3$-term recurrence relation. 
In our case the Gauss elimination method  needs to be applied $3$ times in a row. The details of this calculation are presented in the Appendix B. The final result is the $3$-term recurrence relation
\begin{eqnarray}    \label{3TermFinal}
A_n^{(3)} a_{n+1} + B_n^{(3)} a_n +  C_n^{(3)} a_{n-1} &=& 0, \nonumber \\
A_0^{(3)} a_{1} + B_0^{(3)} a_0   &=& 0  .
\end{eqnarray}
The coefficients $A_n^{(3)}, B_n^{(3)}, C_n^{(3)}$ are functions of the original coefficients $A_n, \dots, F_n$. The explicit dependence is calculated in the Appendix B, see (\ref{contfr3termcoeff}) and (\ref{contfr3termcoeff1}).

In order to solve the $3$-term recurrence relation (\ref{3TermFinal}), we start with the following observation. Since $r_+$ is a regular
singular point, the general expansion (\ref{generalpowersolution}) converges for $r_+   \le r < \infty$. Demanding convergence at  $ r = \infty $ implies  that the  sum $\sum_n a_n$ also converges.
Therefore, defining the quantity $R_n$ by  $R_n = - \frac{a_{n+1}}{a_n},$ we see that the
infinite series (\ref{generalpowersolution}) will converge if $R_n$  decreases sufficiently fast with the increase of $n$. Moreover, from the first relation in (\ref{3TermFinal}) we see that the coefficients $a_n$  must satisfy the recurrence relation
\begin{equation}
R_{n-1} = \frac{C^{(3)}_n}{B^{(3)}_n - A^{(3)}_n R_n}, \quad  \text{that is}  \quad
\frac{a_n}{a_{n-1}} =  \cfrac{-C^{(3)}_n}{B^{(3)}_n - \cfrac{A^{(3)}_n C^{(3)}_{n+1}}{B^{(3)}_{n+1} - \cfrac{A^{(3)}_{n+1} C^{(3)}_{n+2}}{B^{(3)}_{n+2} - \cdot \cdot \cdot}}} .
\end{equation}
This relation, in combination with the second relation in (\ref{3TermFinal}) leads to the  infinite continued  fraction equation
\begin{equation}   \label{infcontfrac}
0=B^{(3)}_0 - \cfrac{A^{(3)}_0 C^{(3)}_1}{B^{(3)}_1 - \cfrac{A^{(3)}_1 C^{(3)}_2}{B^{(3)}_2 -\cfrac{A^{(3)}_2 C^{(3)}_3}{ B^{(3)}_3 - \cdot \cdot \cdot  \cfrac{A^{(3)}_n C^{(3)}_{n+1}}{B^{(3)}_{n+1} - \cdot \cdot \cdot}  }}}.
\end{equation}
The quantity  $R_n$ may be interpreted as being the remaining part of the infinite continued fraction. It  is supposed to decrease relatively fast as $n$ grows.
Hence, the convergence of the series (\ref{generalpowersolution}) is ensured if the
coefficients $a_n$  satisfy the equation  (\ref{infcontfrac}) and $R_n$ decreases with increasing $n$.
The solution to this infinite continued fraction equation gives
the QNM frequencies. The continued fraction relation (\ref{infcontfrac})
can be inverted any number of times. Numerically, the $n$-th
QNM frequency is defined to be the most stable root of the
$n$-th inversion of the continued fraction relation (\ref{infcontfrac}),
\begin{equation}   \label{infcontfrachigher}
B^{(3)}_n - \cfrac{A^{(3)}_{n-1} C^{(3)}_n}{B^{(3)}_{n-1} - \cfrac{A^{(3)}_{n-2} C^{(3)}_{n-1}}{B^{(3)}_{n-2} - \cdot \cdot \cdot  \cfrac{A^{(3)}_0 C^{(3)}_1}{ B^{(3)}_0  }}} = \cfrac{A^{(3)}_{n} C^{(3)}_{n+1}}{B^{(3)}_{n+1} - \cfrac{A^{(3)}_{n+1} C^{(3)}_{n+2}}{B^{(3)}_{n+2} - \cdot \cdot \cdot  }}, \qquad   n=1,2,3\dots .
\end{equation}
The fundamental mode is obtained as the most stable root of the equation (\ref{infcontfrac}).

\subsection{Numerical results}

Now we use the root finding algorithm (Wolfram Mathematica) to determine the QNM spectrum
from equations (\ref{infcontfrac}) and (\ref{infcontfrachigher}). The NC parameter $a$ is
fixed to $a=0.01$. We present our results graphically.

The dependence of the fundamental QNM frequency $\omega = {\mbox Re}\, \omega +i {\mbox Im}\,
\omega$ on the charge\footnote{The charge of the RN black hole $Q$ is fixed.} of the scalar field
$qQ$  is shown in Figure 3. The remaining parameters are fixed as follows: $\mu =0.05$, $l=1$ and
$M=1$ in accordance with \cite{QNMRNBrazilci, Chowdhury:2018izv}. The
extremality of the RN black hole is controlled by the value of $\frac{Q}{M}$. We present results for
ten different cases, with $\frac{Q}{M}$ varying from $0.01$ to $0.999999$.
\begin{center}
\begin{tabular}{lll}
\includegraphics[scale=0.33]{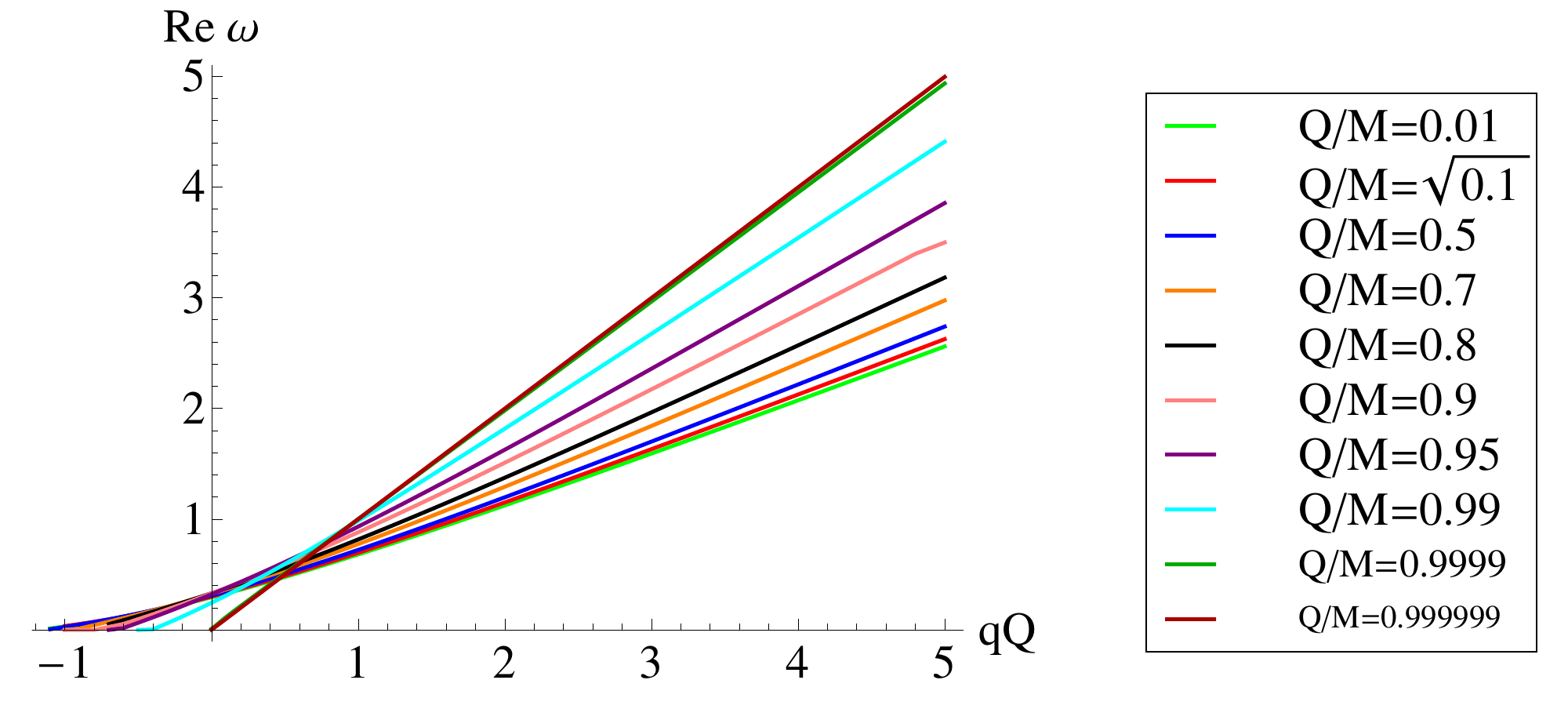}& & \hspace{-3mm}\includegraphics[scale=0.33]{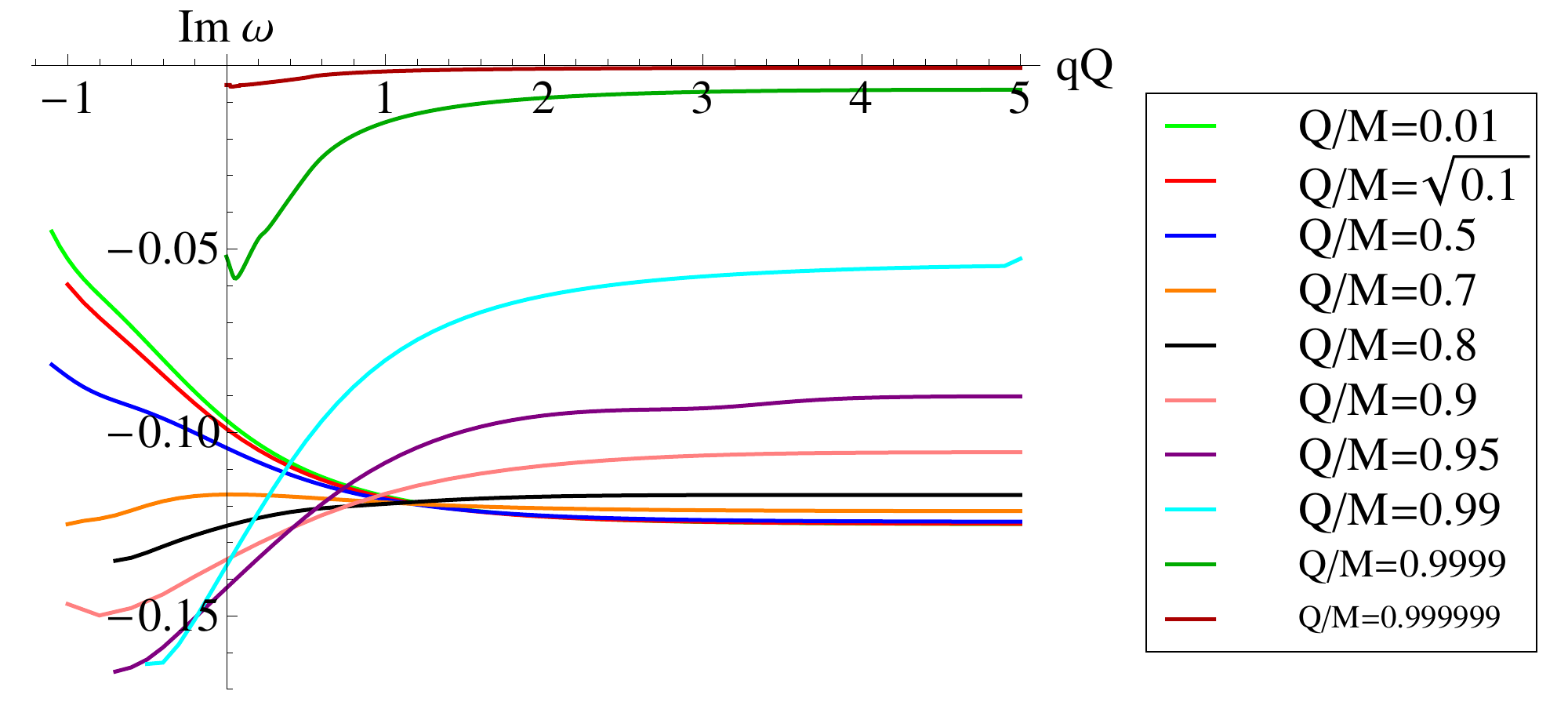}\\
\multicolumn{3}{l}{{\scriptsize {\bf Figure 3:} Dependence of Re\,$\omega$ (left) and Im\,$\omega$ (right) on the charge $qQ$ of the scalar field with the mass }}\\ 
\multicolumn{3}{l}{{\scriptsize $\mu=0.05$ and the orbital momentum $l=1$. Different extremalities are shown in different colors.}}\end{tabular} 
\end{center}
One can see immediately from Figure 3 that, as the electromagnetic interaction increases ($qQ$ increases), the imaginary part of the quasinormal frequency approaches a constant value, while the real part seems to grow linearly. In addition, as $qQ$  decreases, the real part of the fundamental frequency Re\,$\omega$ approaches zero at some critical value of the electromagnetic
coupling. This behavior is similar to the behavior in the commutative case \cite{QNMRNBrazilci}. As this critical value is approached, the continued fraction method seems to converge slower.

The imaginary part of the fundamental frequency ${\mbox Im}\, \omega$ changes its behavior as the extremal limit is approached, at some point between $\frac{Q}{M}=0.8$ and $\frac{Q}{M}=0.9$.  As the extremal limit $\frac{Q}{M}=1$ is approached, the imaginary part of the fundamental frequency becomes smaller, while the real part approaches $\frac{qQ}{r_+} \approx q$.

The existence of the modes with arbitrarily small
imaginary part in the near-extremal limit could be an artifact of the complicated
continued fraction equation, as suggested in \cite{QNMRNBrazilci}. Namely, Leaver’s original method fails to converge for extremal black holes. More precisely, when $\frac{Q}{M}\to 1$, the regular singularities at $r =r_+$ and $r=r_-$ merge at $r=M$, becoming an irregular singularity. Therefore, it is not expected that a power series expansion around $r=r_+$ will have a nonzero
radius of convergence. Onozawa et al. \cite{Onozawa} have proposed a modification of Leaver’s method to deal with such type of equations and have successfully applied it to uncharged fields around an extremal RN black hole. The obtained results are in very good agreement with the ones obtained
for nearly extremal black holes using Leaver’s original method. In our previous work \cite{Ciric:2017rnf}, we treated equation (\ref{EoMR}) analytically in the near extremal limit and we verified the existence of the modes
with arbitrary small imaginary parts. In fact, the QNMs obtained in \cite{Ciric:2017rnf} in the near extremal limit are in excellent agreement with
the ones obtained through the continued fraction method in the limit $\frac{Q}{M}\to 1$. 

\begin{center}
\begin{tabular}{lll}
\includegraphics[scale=0.33]{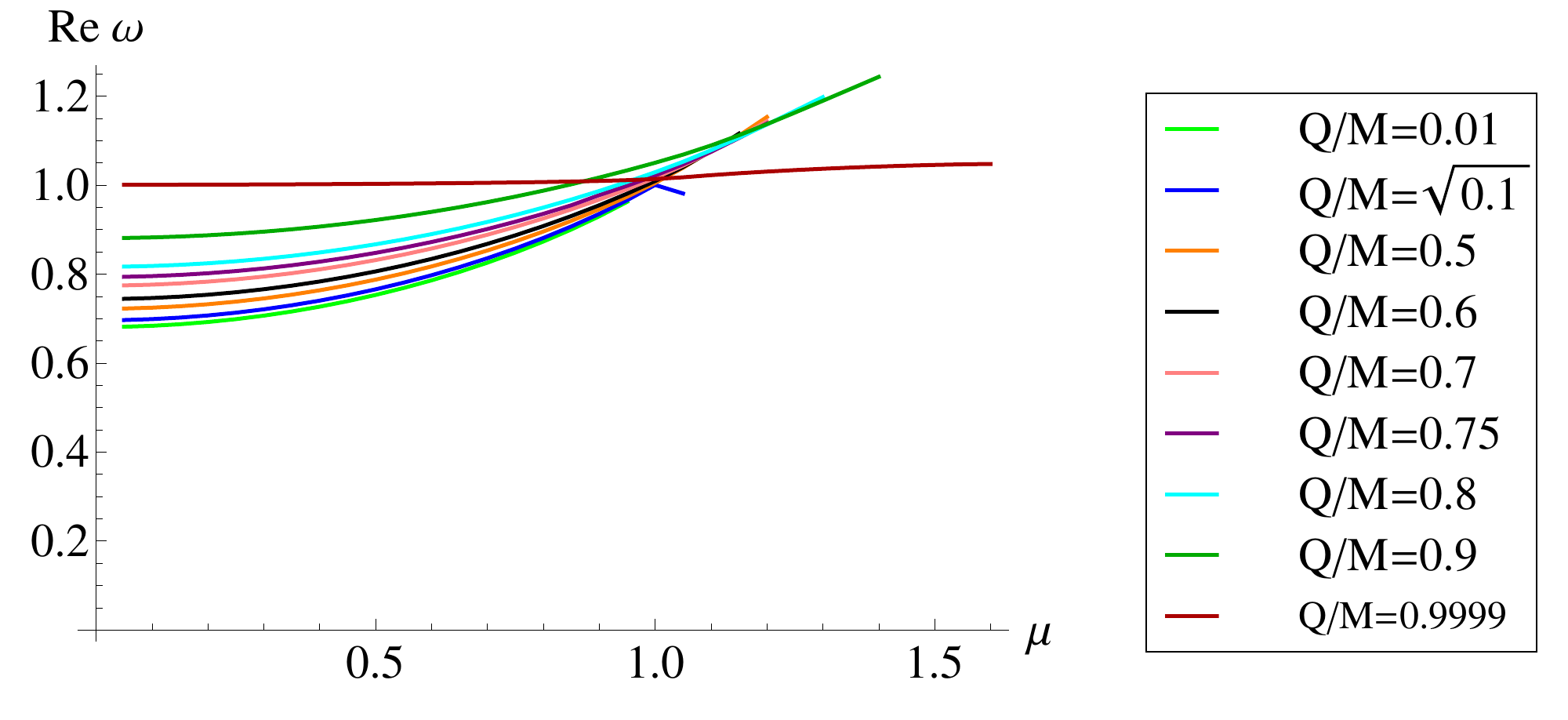}& & \hspace{-3mm}\includegraphics[scale=0.33]{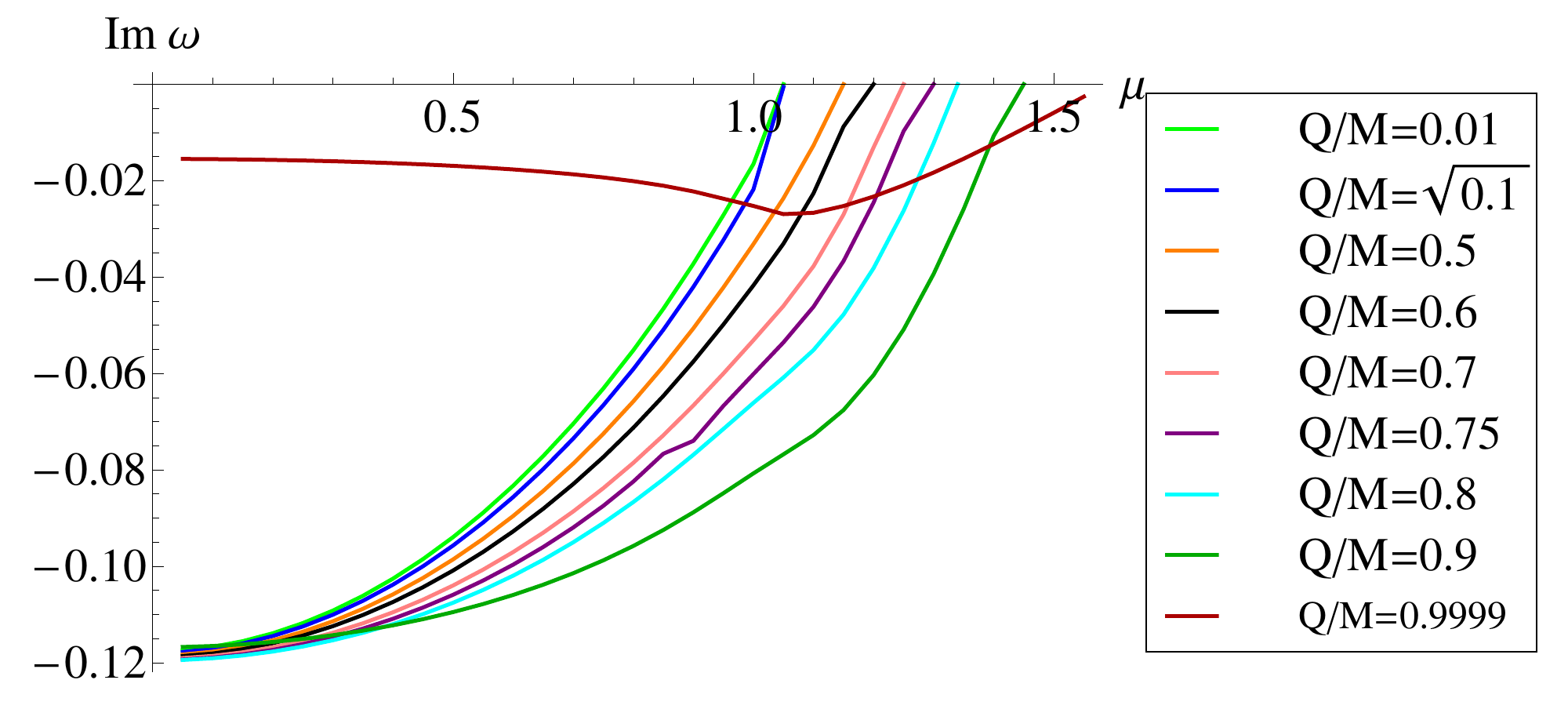}\\
\multicolumn{3}{l}{{\scriptsize {\bf Figure 4:} Dependence of Re\,$\omega$ (left) and Im\,$\omega$ (right) on the mass $\mu$ of the scalar field with the charge }}\\ 
\multicolumn{3}{l}{{\scriptsize $qQ=1$ and the orbital momentum $l=1$.}}
\end{tabular} 
\end{center} 
In Figure 4 we show the dependence of the fundamental frequency Re\,$\omega$  and Im\,$\omega$ on
the mass $\mu$ of the scalar field $\phi$. The remaining parameters are fixed to:
$qQ=1$, $l=1$ and $M=1$. Different extremalities are shown in different colors.
The value of $\frac{Q}{M}$ varies from $0.01$ up to $0.9999$. The existence of quasi-resonances,
special values of the mass when ${\mbox Im}\,\omega =0$, is clearly visible for all extremalities
presented. The existence of quasi-resonances is in the agreement with the results obtained in the
near-extremal approximation in \cite{Ciric:2017rnf}.

Finally, let us comment on the effects of noncommutative deformation in our model. It is clear from
equations (\ref{6contfr}) and (\ref{contfr1}) that the QNM frequencies depend on the value of $am$,
where $a$ controls the NC deformation and $m = -l,\ -l +1,..., l$ is the projection of the orbital
momentum $l$. Therefore, we expect to find a frequency splitting for different values of $m$. To
make this observation more explicit, in Figures 5 and 6 we show the frequency splitting as a
function of the scalar field charge $q$ (Figure 5) and of the scalar field mass $\mu$ (Figure 6).
The splitting can be graphically presented for an arbitrary value of $\frac{Q}{M}$, taking into
account comments on the near extremal limit from the previous paragrph. In the example shown in
Figures 5 and 6 we fixed the extremality to $\frac{Q}{M}=0.5$, that is we show splitting in the
non-extremal case. The splitting is defined as $\omega^\pm = \omega (m=\pm 1) -\omega (m=0)$.

The relative frequency splitting can be estimated from Figures 3 and 5 and Figures 4 and 6. For
example, in the case of fixed mass $\mu=0.05$ and $\frac{Q}{M}=0.5$ we find $\delta_{\mbox {\tiny
Re}} \sim \frac{{\mbox Re}\,\omega^+}{{\mbox Re}\,\omega} \sim 10^{-4}$ and $\delta_{\mbox {\tiny
Im}} \sim \frac{{\mbox Im}\,\omega^+}{{\mbox Im}\,\omega} \sim 10^{-4}$. One can check that in the
case of fixed charge $qQ=1$, the splitting is approximately of the same order.

\begin{center}
\begin{tabular}{lll}
\includegraphics[scale=0.33]{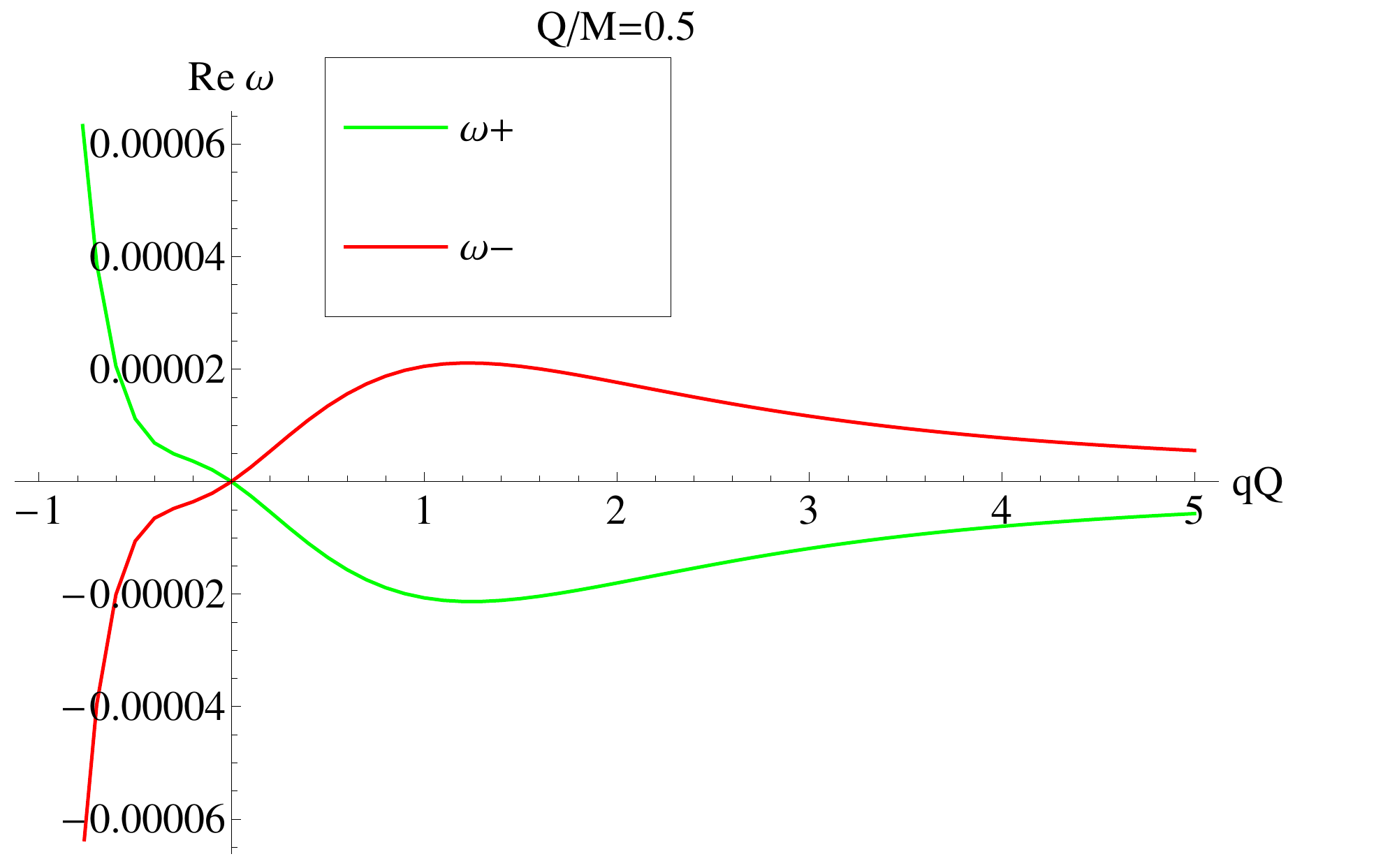}& & \hspace{-3mm}\includegraphics[scale=0.33]{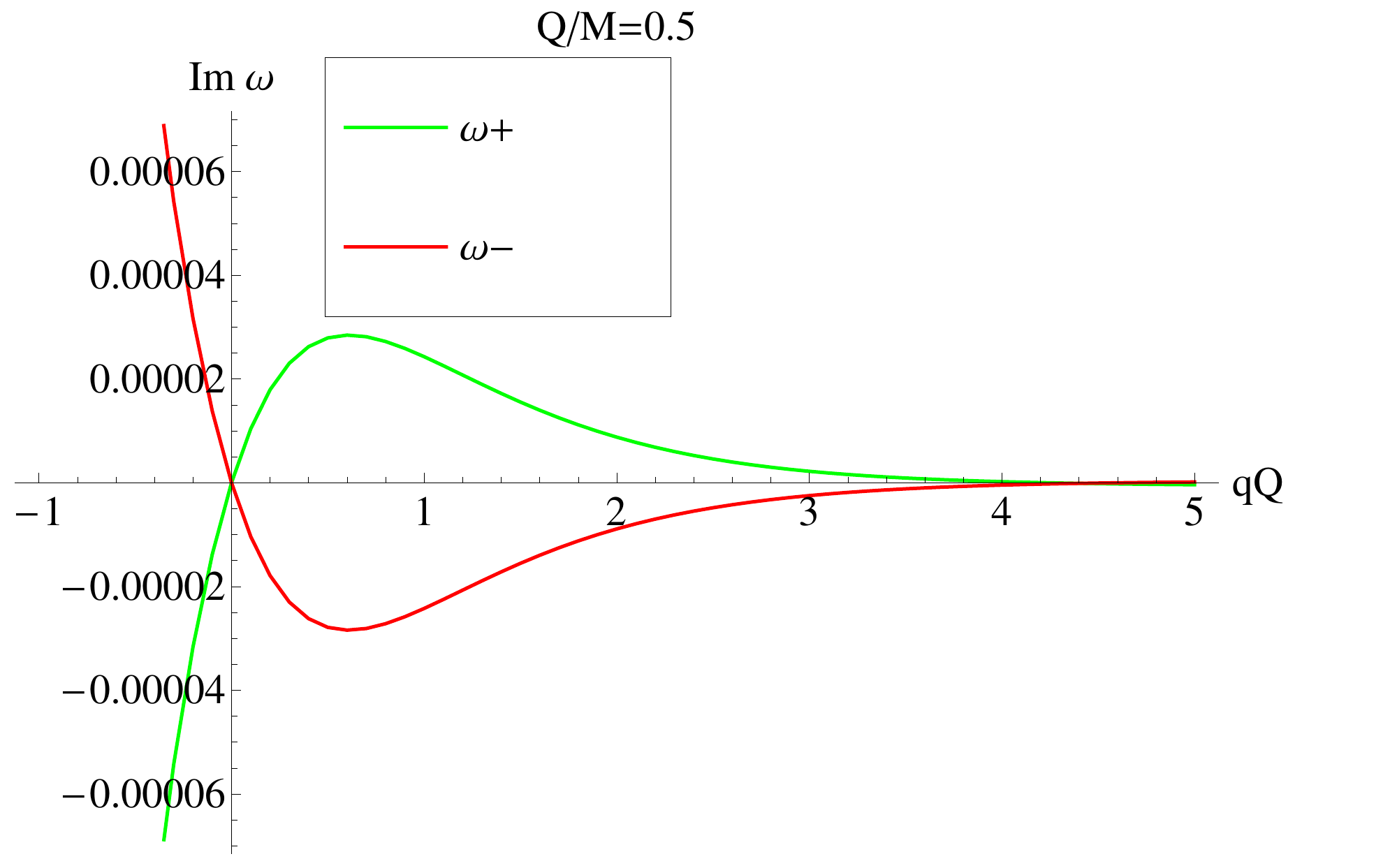}\\
\multicolumn{3}{l}{{\scriptsize {\bf Figure 5:} Dependence of Re\,$\omega^\pm$ (left) and Im\,$\omega^\pm$ (right) on the charge $qQ$ of the scalar field with the}}\\ 
\multicolumn{3}{l}{{\scriptsize mass $\mu=0.05$, orbital momentum $l=1$ and extremality $\frac{Q}{M}=0.5$.}}
\end{tabular} 
\end{center}
\begin{center}
\begin{tabular}{lll}
\includegraphics[scale=0.33]{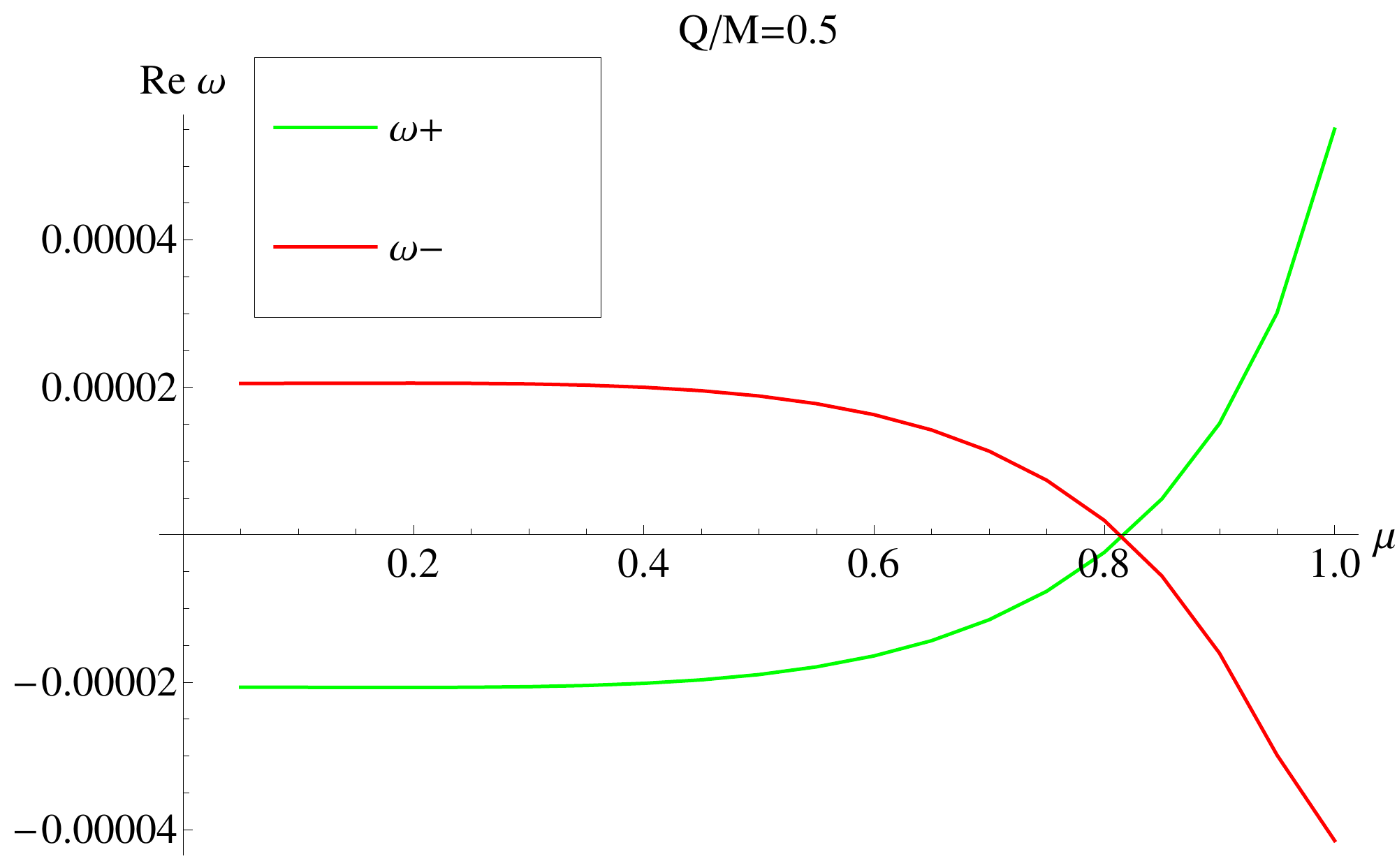}& & \hspace{-3mm}\includegraphics[scale=0.33]{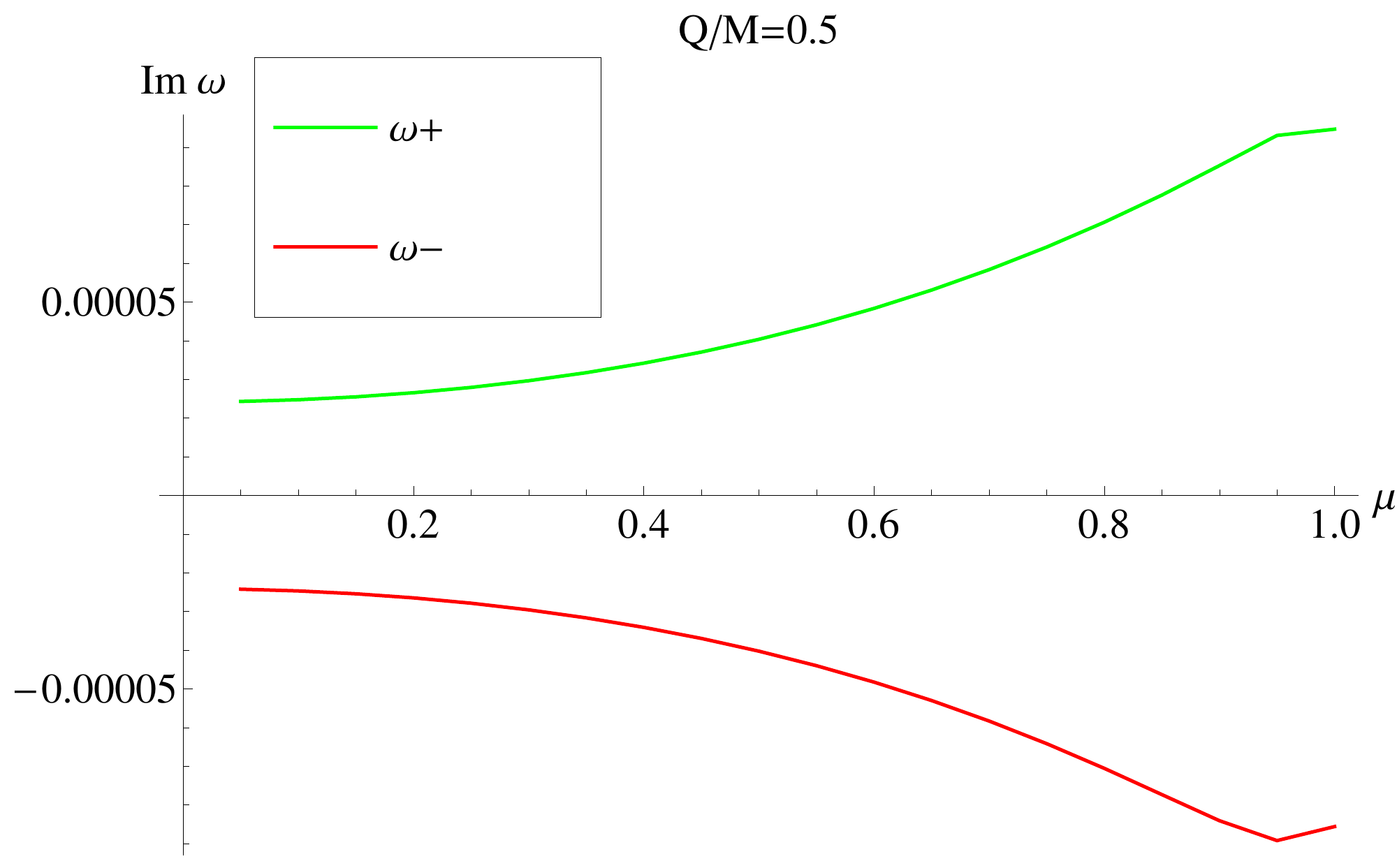}\\
\multicolumn{3}{l}{{\scriptsize {\bf Figure 6:} Dependence of Re\,$\omega^\pm$ (left) and Im\,$\omega^\pm$ (right) on the mass $\mu$ of the scalar field with the charge }}\\ 
\multicolumn{3}{l}{{\scriptsize $qQ=1$, orbital momentum $l=1$ and extremality
$\frac{Q}{M}=0.5$.}}
\end{tabular} 
\end{center}

It is important to stress that in the limit $a \rightarrow 0$ (and $\mu \rightarrow 0$) our results
reduce to the results presented in \cite{QNMRNBrazilci, Chowdhury:2018izv, Konoplya:2013rxa},
meaning that we have been able to reproduce  in full detail the same  graph profiles as in these
references. In particular, for the uncharged massive scalar field in the absence of
noncommutativity and for  $\frac{Q}{M} =\sqrt{0.1} \approx$  0.316  we checked that the curves
showing the dependence of Re\,$\omega$ and Im\,$\omega$ on the mass $\mu$ are in a perfect
agreement with the corresponding curves in reference \cite{Chowdhury:2018izv}, see the graphs on
Figures 2c) and 2d) respectively, specified for the case of the scalar charge $ s =  0.1$. In
Figures 3 and 4 the dark blue lines correspond to this particular extremality $\frac{Q}{M} =0.316$.
However, in this case, unlike in \cite{Chowdhury:2018izv} the scalar field is charged and in
addition there are effects of noncommutativity. Another example which may be interesting to
confront with our results
is the fundamental mode of the massless scalar field in a non-rotational Kerr-Newman  background
which was studied in reference \cite{Konoplya:2013rxa}, see Figure 3 in \cite{Konoplya:2013rxa} for
the case $r_- = 0.95r_+$. This case roughly corresponds to the extremality ratio of $Q/M =0.9999$,
which is represented on our Figure 3 by the dark green line. A close inspection of two graph
profiles  shows a high level of agreement. In particular, the imaginary parts in both cases saturate
roughly at the same value and both have the minimum lying on the vertical  axis. The  small
discrepancy  between the two profiles may be attributed to the fact that the mass of the scalar
probe in these two cases is not the same,  as well as to a small effect caused by noncommutativity.

\section{Discussion and final remarks}

In this paper we studied the QNM spectrum of the NC scalar perturbation in the non-extremal RN background in more depth. In our previous work \cite{Ciric:2017rnf} we did the analytic analysis limited to the near-extremal geometry. Here we overcome this limitation by using the WKB and the continued fraction methods. The WKB method is well defined for large values of the orbital momentum $l$, and due to our additional approximations is limited to the parameter region $l+1 \ll qQ$. In this approximation we solved the WKB condition analytically and plotted our results in Figures 1 and 2. If  one makes the additional approximation $l\ll qQ\ll l^2,$ the massless, commutative limit of our results (\ref{wkbQNMcondition4}) and (\ref{wkbQNMcondition5}) agrees with the results obtained in \cite{HodWKBRN}. 

Looking for a less restrictive method, we then turned to the continued fraction method. To our
knowledge, this is the first time the continued fraction method is applied to the noncommutative QNM
spectrum problem. The NC deformation induces the 6-term recurrence relation (\ref{6contfr}) and we
use the Gauss elimination procedure to reduce it to the 3-term recurrence relation
(\ref{3TermFinal}). We used the root finding algorithm to solve the 3-term recurrence relation and
obtain the fundamental QNM frequency. The results we obtain are plotted in Figures 3-6 and are in
good qualitative agreement with our previous results in \cite{Ciric:2017rnf}. We discuss some of the
properties of the QNM spectrum in Section 4.2. Here, for completeness, we also show the results for
$l=2$. In Figure 7 we plot the dependence of the fundamental frequency $\omega$ on the scalar field
charge $q$, while in Figure 8 we plot the corresponding frequency splitting defined as
$\omega^{\pm \pm} = \omega (m=\pm 2) -\omega (m=0)$.

\vspace{3mm}
\begin{center}
\begin{tabular}{lll}
\includegraphics[scale=0.33]{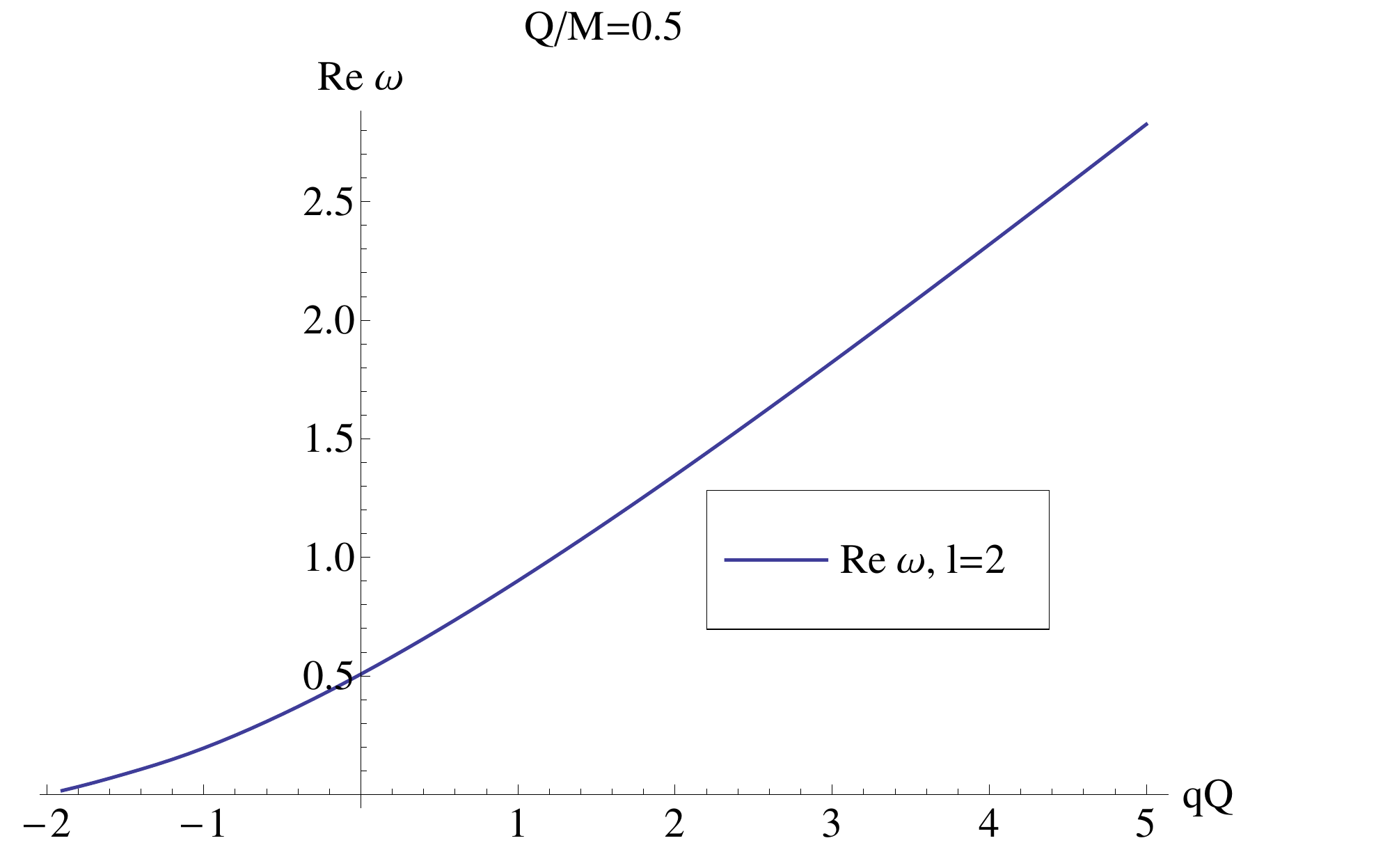}& & \hspace{-3mm}\includegraphics[scale=0.33]{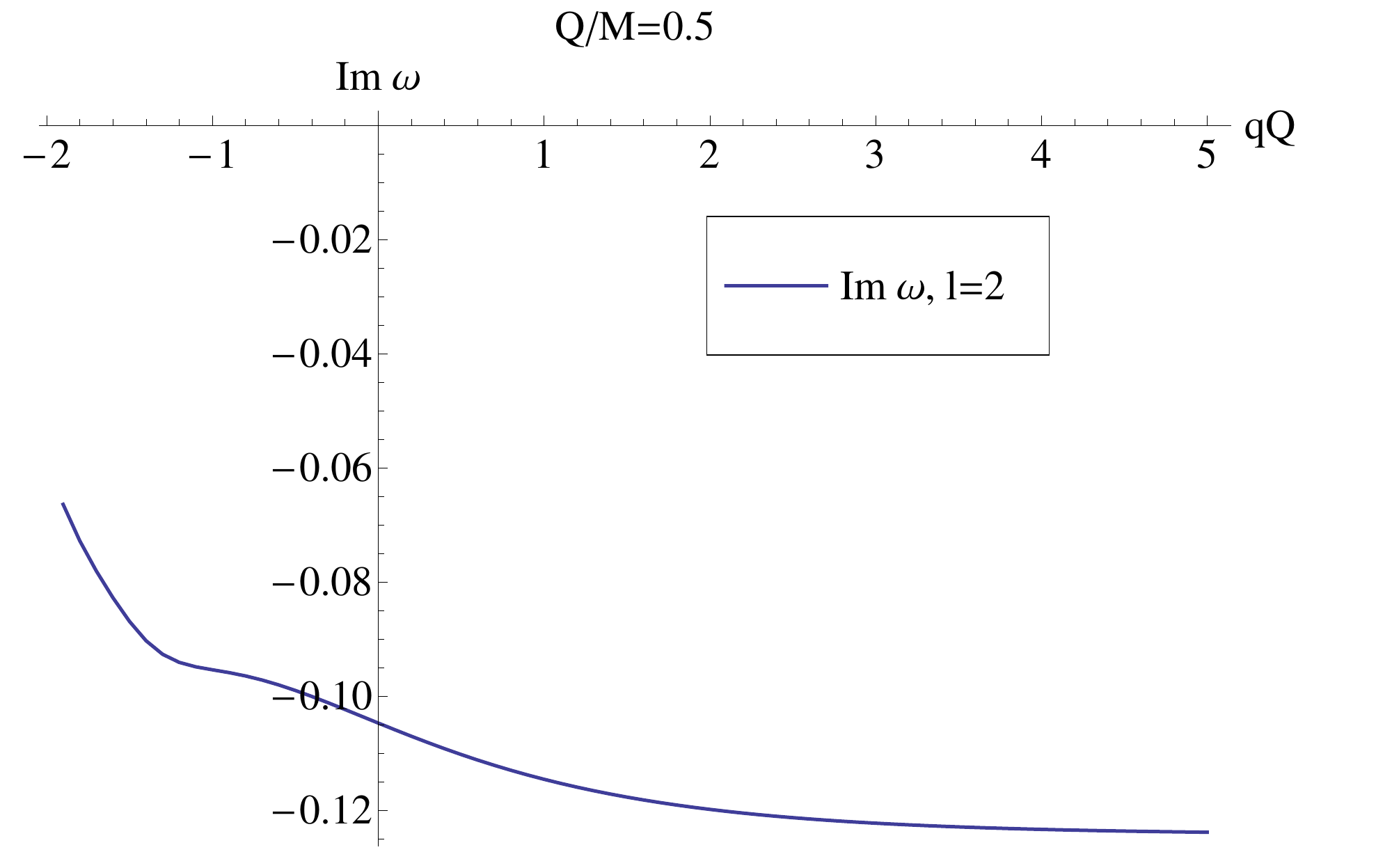}\\
\multicolumn{3}{l}{{\scriptsize {\bf Figure 7:} Dependence of Re\,$\omega$ (left) and Im\,$\omega$ (right) on the charge $qQ$ of the scalar field with the mass }}\\ 
\multicolumn{3}{l}{{\scriptsize $\mu=0.05$, the orbital momentum $l=2$ and the extremality
$\frac{Q}{M}=0.5$.}}\\
\end{tabular} 
\end{center}

\begin{center}
\begin{tabular}{lll}
\includegraphics[scale=0.33]{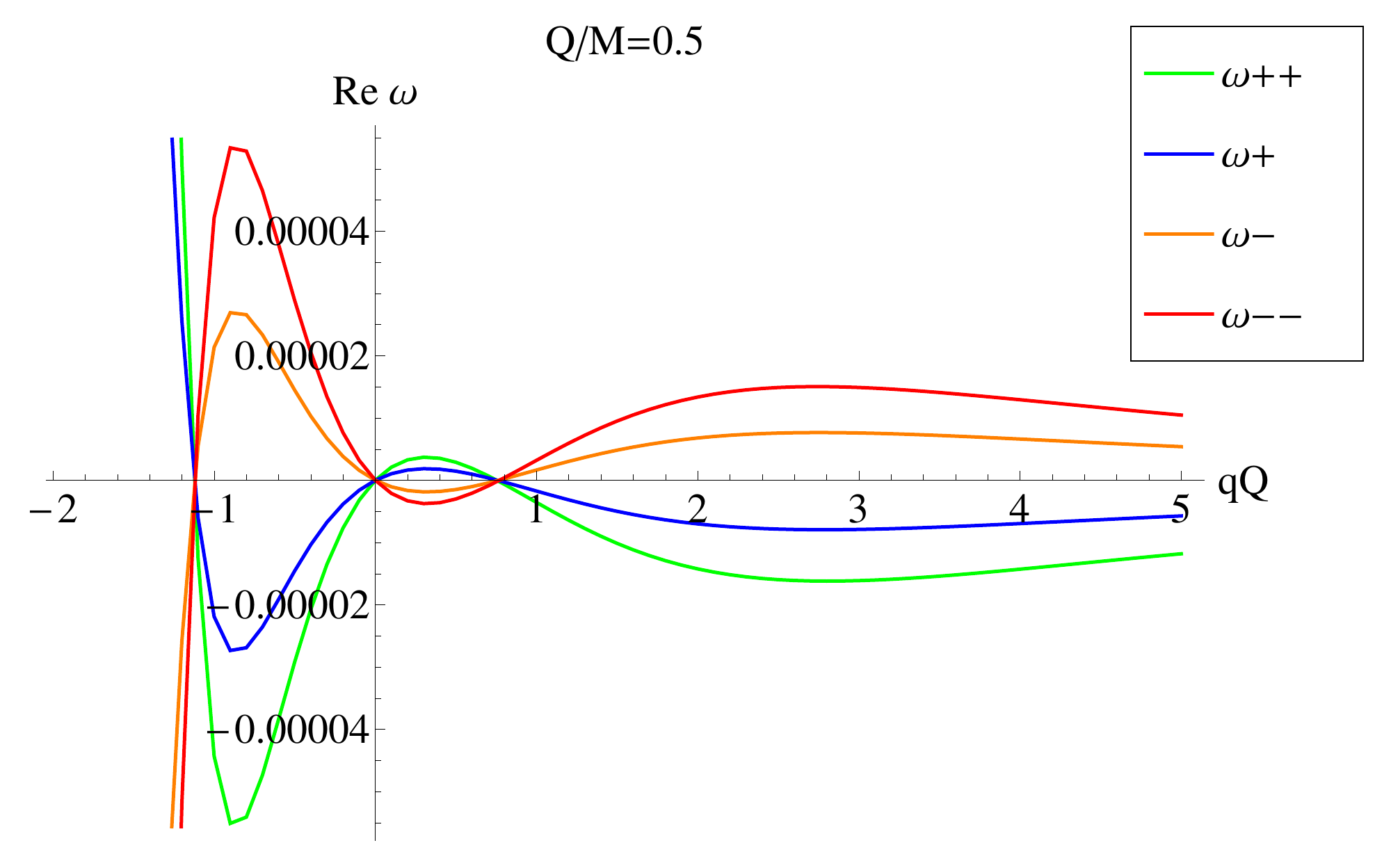}& & \hspace{-3mm}\includegraphics[scale=0.33]{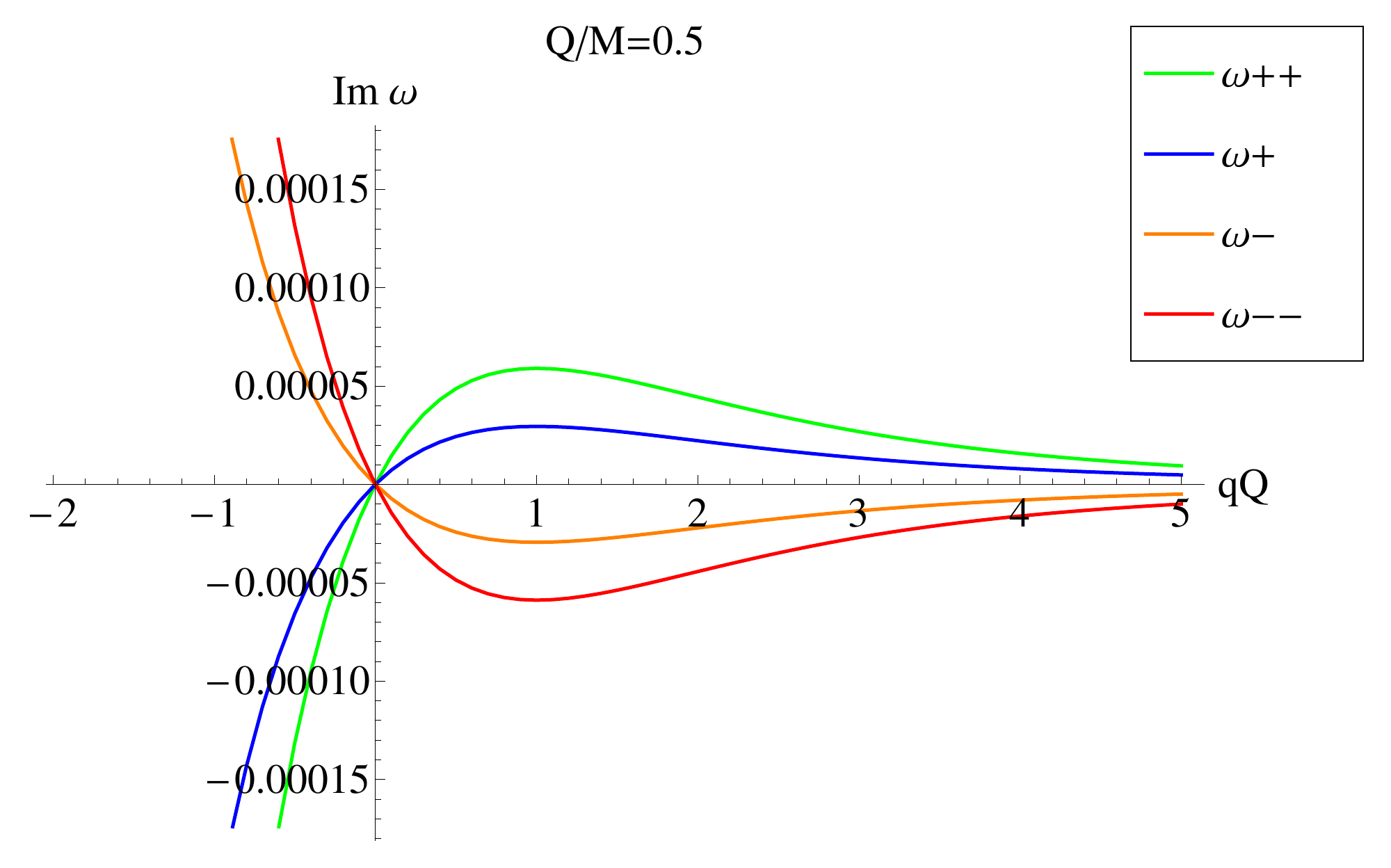}\\
\multicolumn{3}{l}{{\scriptsize {\bf Figure 8:} Dependence of Re\,$\omega^\pm $, Re\,$\omega^{\pm
\pm} $ (left) and Im\,$\omega^\pm$, Im\,$\omega^{\pm \pm}$ (right) on the charge $qQ$ of the scalar
}}\\ 
\multicolumn{3}{l}{{\scriptsize field with the mass $\mu=0.05$, orbital momentum $l=2$ and
extremality $\frac{Q}{M}=0.5$.}}\\
\end{tabular} 
\end{center}

Finally, to compare the results obtained by the continued fraction method with those obtained in 
the WKB approximation, we present on Figure 9 the real and the imaginary part of the fundamental
frequency $\omega$ in the case of $l=100$ and the NC parameter fixed at $a=0.000001$.
\begin{center}
\begin{tabular}{lll}
\includegraphics[scale=0.33]{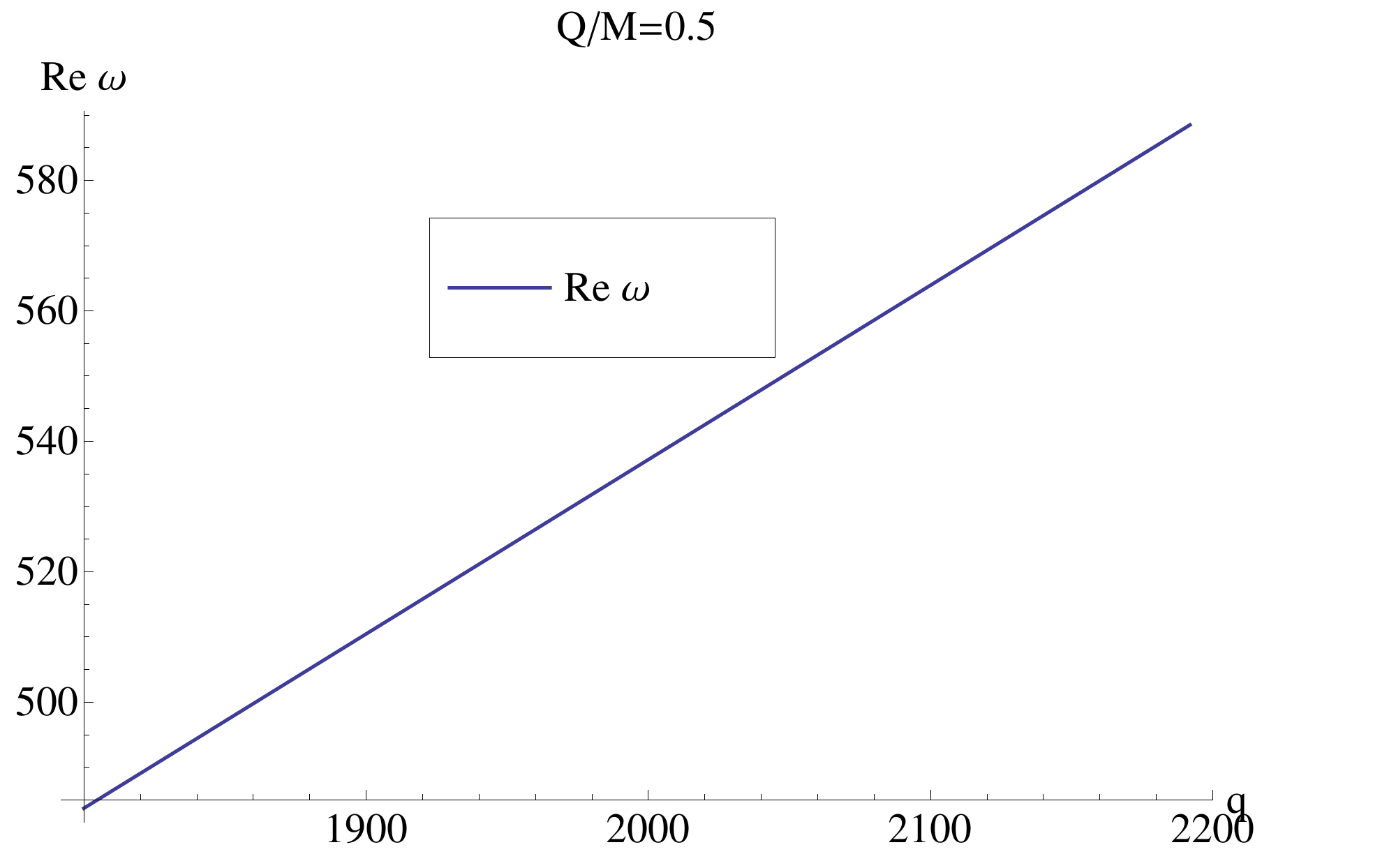}& & \hspace{-3mm}\includegraphics[scale=0.33]{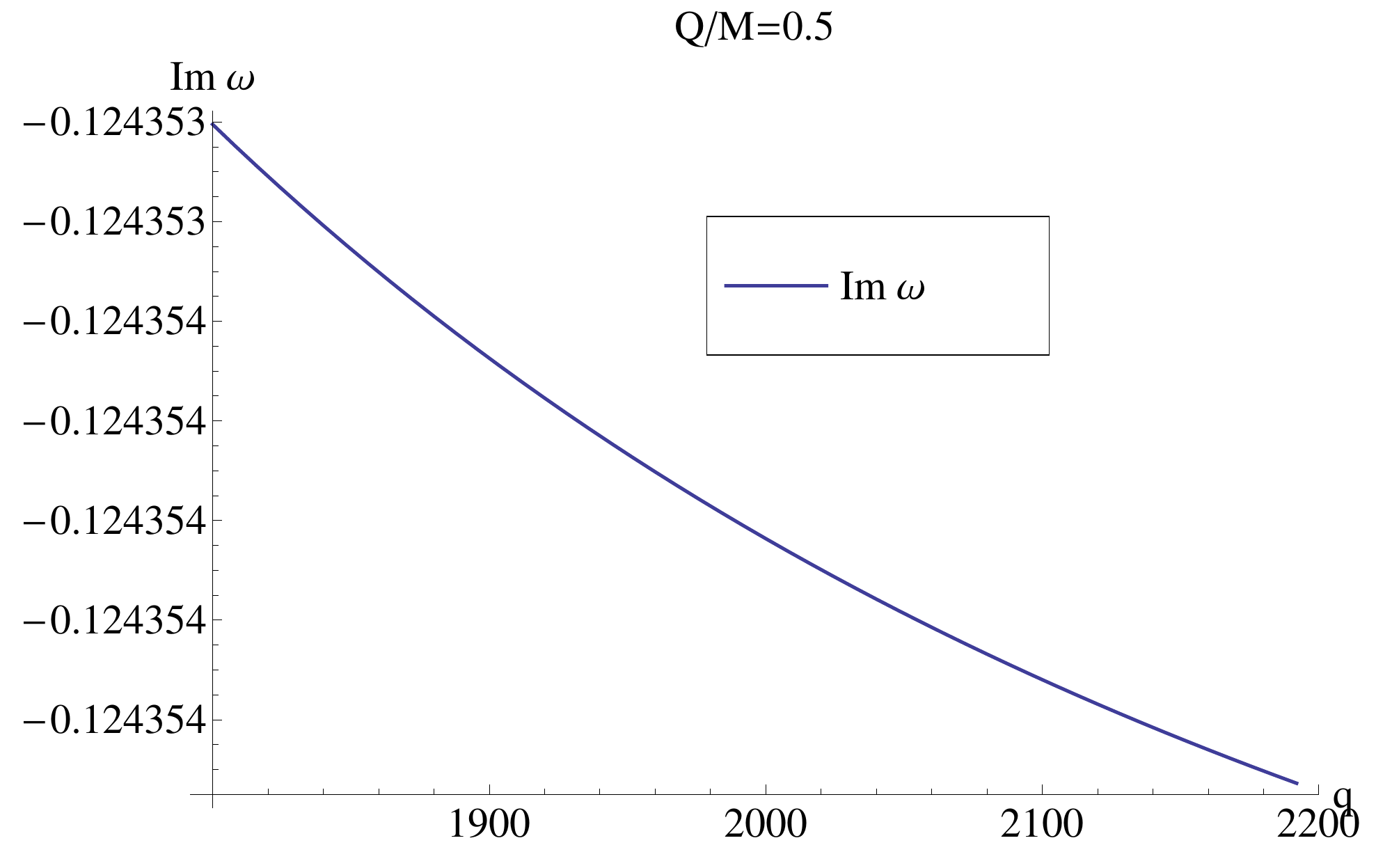}\\
\multicolumn{3}{l}{{\scriptsize {\bf Figure 9:} Dependence of Re\,$\omega$ (left) and Im\,$\omega$ (right) on the charge $qQ$ of the scalar field with the mass }}\\ 
\multicolumn{3}{l}{{\scriptsize $\mu=0.05$ and the angular momentum $l=100$.}}
\end{tabular} 
\end{center}
The frequency splittings are presented in Figure 10.
\begin{center}
\begin{tabular}{lll}
\includegraphics[scale=0.33]{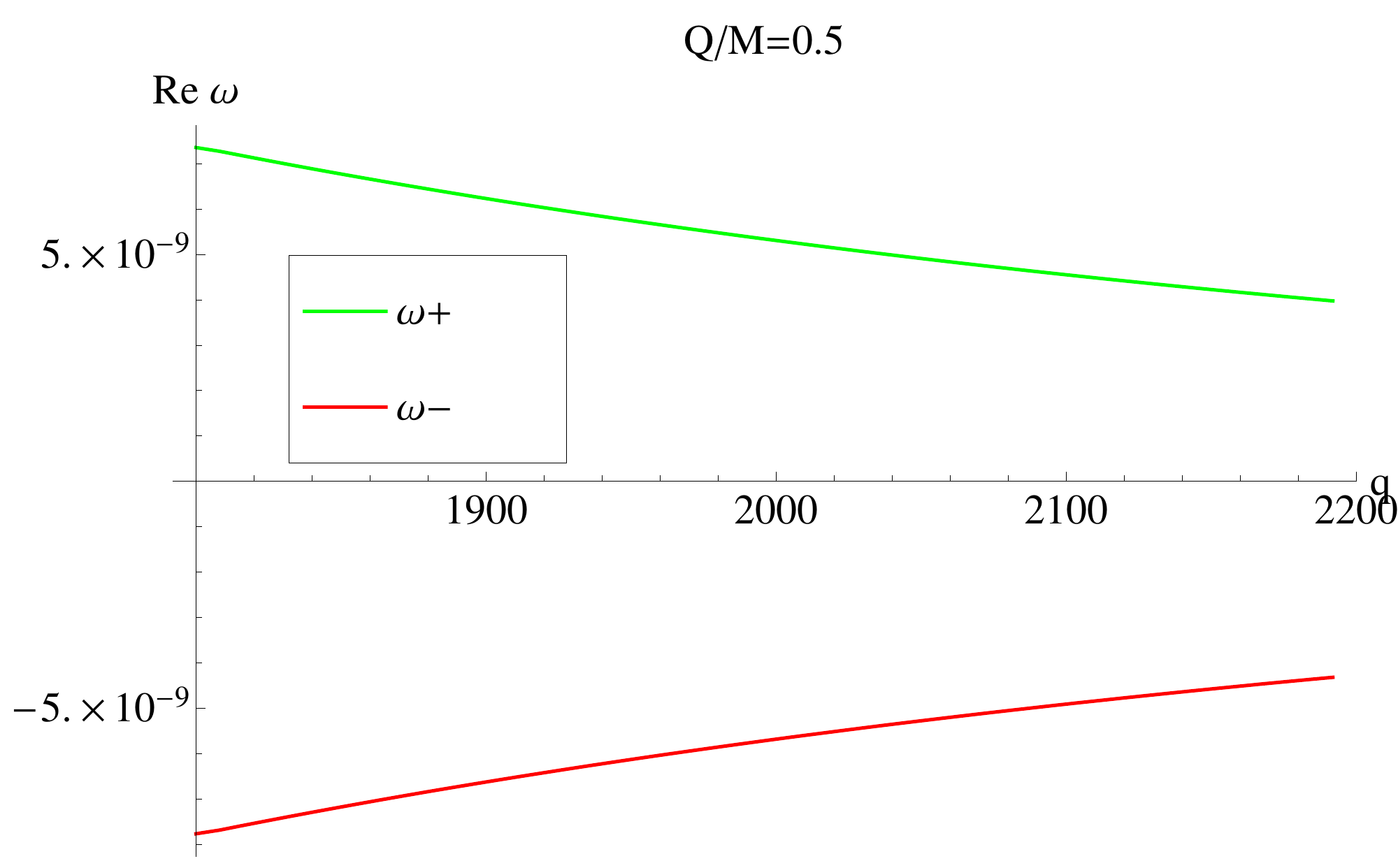}& & \hspace{-3mm}\includegraphics[scale=0.33]{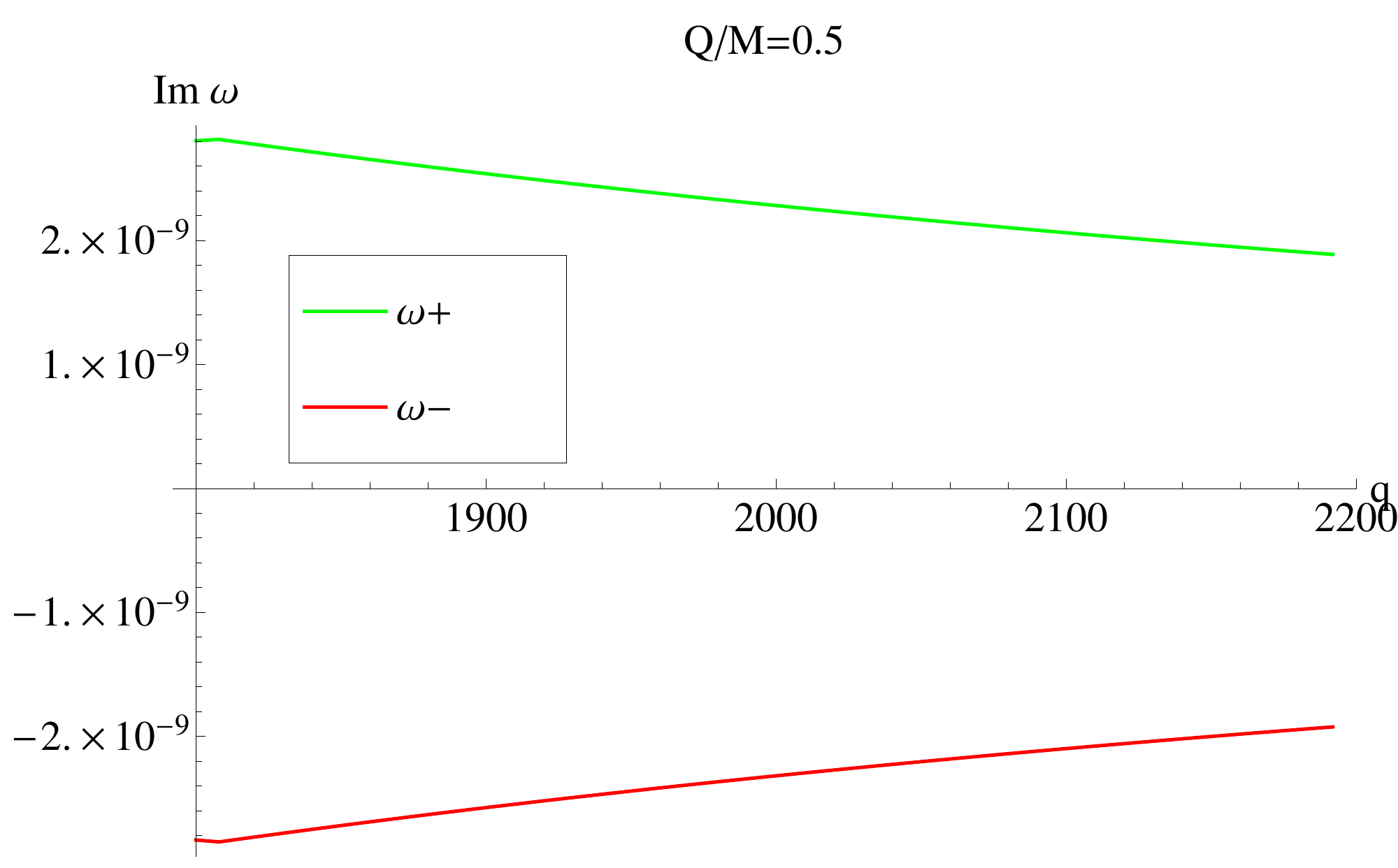}\\
\multicolumn{3}{l}{{\scriptsize {\bf Figure 10:} Dependence of Re\,$\omega^\pm$ (left) and
Im\,$\omega^\pm$ (right) on the charge $qQ$ of the scalar field with the }}\\ 
\multicolumn{3}{l}{{\scriptsize mass $\mu=0.05$ and the angular momentum $l=100$.}}
\end{tabular} 
\end{center}
Comparing with Figures 1-2, we notice a qualitative agreement between these two methods. Quantitative results are a bit different, but that was expected. WKB is an approximative method and we made additional approximations in order to obtain an analytic result for the frequencies (\ref{wkbQNMcondition5}).

As we mentioned before, this is the first time the continued fraction method is applied to a NC QNM spectrum problem. In our future work, we plan to investigate spinor and vector NC QNM spectrum. Of course, we are interested to go beyond the semiclassical analysis used in this paper. To do that, we need a full NC gravity action. We hope that we can make some progress in this direction using the model developed in \cite{EPL2017}.

\vskip1cm \noindent 
{\bf Acknowledgement}
We would like to thank Mauricio Richartz, Tajron Juri\'c and Svetislav Mijatovi\' c for
fruitful discussion and useful comments. The work of M.D.C. and N.K. is
supported by project
ON171031 of the Serbian Ministry of Education and Science. The work of A.S. is  partially supported by the
H2020 CSA Twinning project No. 692194, RBI-T-WINNING as well as by the project "Synergy to Success: RBI-T-WINNING and ESIF Associated in Strengthening the Excellence of the Institute of Theoretical Physics of the Rudjer Boskovic Institute (RBI-TWINN-SIN)" . This work is partially
supported by ICTP-SEENET-MTP Project NT-03 "Cosmology-Classical and Quantum Challenges" in frame of the Southeastern European Network in
Theoretical and Mathematical Physics and by the Action MP1405 QSPACE from the European 
Cooperation in Science and Technology  (COST).

\appendix

\renewcommand{\theequation}{\Alph{section}.\arabic{equation}}
\initiate
\section{Finding the $6$-term recurrence relation}

In this Appendix we derive the $6$-term recurrence relations  (\ref{6contfr}) and find the explicit expressions for the coefficients $A_n, B_n, C_n, D_n, E_n, F_n$ that appear there. Starting with $R(r) = e^{i \Omega r} \psi (r)$, equation (\ref{EoMR}) reduces to
\begin{align}
&   (r-r_+)(r-r_-)  \frac{\d^2 \psi}{\d r^2}  \nn\\
& + \Big[  2(r-M) + 2i \Omega (r- r_+)(r-r_-) - iam qQ  \frac{(r-r_+)(r-r_-)}{r^2}  \Big]  \frac{\d \psi}{\d r}  \nn \\
&  + \Big[ 2(r -M) i\Omega  - \Omega^2 (r-r_+)(r-r_-) - l(l+1) + \frac{r^4}{(r-r_+)(r-r_-)} {\Big( \omega - \frac{qQ}{r}   \Big)}^2 \nn \\
&   - \mu^2 r^2  - iam qQ  \Big( \frac{M}{r^2}  - \frac{Q^2}{r^3} \Big) - iamqQ i \Omega \frac{(r-r_+)(r-r_-)}{r^2}   \Big] \psi  =0. \label{derivacije}
\end{align}
Introducing the new variable  $z = \frac{r-r_+}{r-r_-},$  one has
\begin{equation}
r = \frac{r_+ - zr_-}{1-z}, \qquad  r-r_- = \frac{r_+ - r_-}{1-z}, \qquad   r-r_+ =  \frac{z (r_+ - r_-)}{1-z}. \label{App1PomocneFle}
\end{equation}
Noting that (\ref{EoMR}) has an irregular singularity at $r=+\infty$ and three regular singularities at
$r=0, \; r=r_-$ and $r=r_+,$ we can expand the solution in terms of powers series
around $r=r_+$
\begin{equation}
\psi (r) = {(r-r_-)}^{\epsilon} \sum_{n=0}^{\infty} a_n {\Big( \frac{r-r_+}{r-r_-} \Big)}^{n + \delta} = \frac{{(r_+ - r_-)}^{\epsilon}}{{(1-z)}^{\epsilon}}
\sum_{n=0}^{\infty} a_n z^{n + \delta}. \label{appendix0}
\end{equation}
We have to keep in mind that the complete radial solution 
\begin{equation}
  R(r) = e^{i \Omega r}  {(r-r_-)}^{\epsilon} \sum_{n=0}^{\infty} a_n {\Big( \frac{r-r_+}{r-r_-} \Big)}^{n + \delta}
\end{equation}
must satisfy the  boundary conditions (\ref{ncboundaryconditions}) on the horizon and at  far infinity. The change of coordinates from $r$ to $z$ is accompanied with the transformation of derivatives
\begin{equation}
\frac{\d F}{\d r} =  \frac{{(1-z)}^2}{r_+ - r_-}   \frac{\d F}{\d z},  \qquad
\frac{\d^2 F}{\d r^2} =  \frac{{(1-z)}^3}{{(r_+ - r_-)}^2}  \Big[ (1-z) \frac{\d^2 F}{\d z^2} - 2  \frac{\d F}{\d z}   \Big]   .\nonumber
\end{equation}
Inserting further the expansion (\ref{appendix0}) into   (\ref{derivacije}), 
and using the decomposition $r = \frac{r_+ -zr_-}{1-z}  = r_- + \frac{r_+ - r_-}{1-z},$  results  in
\begin{align}
& z{(1-z)}^2 {(r_+ - r_-)}^{\epsilon} \Bigg[ \frac{\epsilon(\epsilon +1)}{{(1-z)}^{\epsilon + 2}}  \sum_{n=0}^{\infty} a_n z^{n + \delta} + \frac{2\epsilon }{{(1-z)}^{\epsilon + 1}} \sum_{n=0}^{\infty} a_n  (n+\delta)  z^{n + \delta -1}   \nonumber \\
& + \frac{1}{{(1-z)}^{\epsilon}} \sum_{n=0}^{\infty} a_n  (n+\delta)  (n+\delta -1) z^{n + \delta -2} \Bigg] \label{appendix} \\
& + \Bigg[  {(1-z)}^2 + 2i \Omega z (r_+ - r_-) -iamqQ \frac{(r_+ - r_-)z {(1-z)}^2 }{{(r_+ - zr_-)}^2} \Bigg] {(r_+ - r_-) }^{\epsilon}\times\nn\\
& \hspace{5mm}\Bigg[ \frac{\epsilon }{{(1-z)}^{\epsilon + 1}}  \sum_{n=0}^{\infty} a_n z^{n + \delta} + \frac{1}{{(1-z)}^{\epsilon }} \sum_{n=0}^{\infty} a_n  (n+\delta)  z^{n + \delta -1}  \Bigg] \nonumber \\
& + \Bigg[ i\Omega (r_+ - r_-) \frac{1+z}{1-z} - \Omega^2 z \frac{{(r_+ - r_-) }^{2}}{{(1-z)}^2} -l(l+1) \nn\\
& + \frac{{(r_+ - z r_-) }^{2}}{ z {(r_+ - r_-)}^2} {\Bigg(  \omega \bigg( r_-  + \frac{r_+ - r_-}{1-z}    \bigg) -qQ    \Bigg)}^2
-\mu^2 {  \bigg( r_-  + \frac{r_+ - r_-}{1-z}    \bigg) }^2  \nn\\
& - iamqQ \bigg( \frac{M}{r^2} - \frac{Q^2}{r^3}  \bigg) + amqQ\Omega  \frac{ {(r_+ - r_-)}^2 z}{{(r_+ - z r_-)}^2} \Bigg]   \frac{{(r_+ - r_-) }^{\epsilon}}{{(1-z) }^{\epsilon}} \sum_{n=0}^{\infty} a_n  z^{n + \delta } =0. \nn
\end{align}
After multiplying the whole relation with ${(1-z)}^{\epsilon},$ only a few  terms with the fractions will remain. All these terms have  either $1-z$ or its powers in the denominator.  We combine them to get a  simplified expression
\begin{align}
& \frac{2\omega^2 r_- (r_+ - r_-)}{1-z}  \sum_{n=0}^{\infty} a_n z^{n+\delta +1} + \Bigg[  2\omega
\frac{(r_+ - r_-)(\omega r_+ - qQ)}{1-z} + 2i \Omega \frac{(r_+ - r_-) \epsilon z}{1-z}  \nonumber  
\end{align}
\begin{align}
&  + i \Omega (r_+ - r_-) \frac{1+z}{1-z} - \Omega^2 {(r_+ - r_-)}^2 \frac{z}{{(1-z)}^{2}}
+  \omega^2   \frac{{(r_+ - r_-) }^{2} z}{{(1-z)}^2}\nn\\
& - \mu^2 \Bigg(  2 \frac{r_- (r_+ - r_-)}{1-z}  + \frac{{(r_+ - r_-) }^{2}}{{(1 - z) }^{2}}    \Bigg)  \Bigg] \sum_{n=0}^{\infty} a_n z^{n+\delta } \label{appendix1}  
\end{align}
\begin{equation}
 =  \sum_{n=0}^{\infty}  \Big[ i \Omega (r_+ - r_-) + 2\omega (\omega r_+ - qQ) (r_+ - r_-) - \mu^2 (r_+ + r_-) (r_+ - r_-)     \Big] a_n z^{n+\delta }.  \nonumber
\end{equation}
We note that the relation $\Omega^2 = \omega^2 - \mu^2 $  plays a crucial role in the above simplification and that the expression (\ref{epsilondelta}) for $\epsilon$ was utilised in the above calculation. Turning back to the equation (\ref{appendix}), we make use of the  simplification  described above and obtain  the following relation
\begin{align}
& \epsilon^2 \sum_{n=0}^{\infty} a_n z^{n+\delta +1}  + \epsilon \sum_{n=0}^{\infty} a_n z^{n+\delta}  + \sum_{n=0}^{\infty} a_n (n + \delta) z^{n+\delta -1}  \nonumber  \\
& + (2\epsilon -2) \sum_{n=0}^{\infty} a_n (n + \delta) z^{n+\delta }   \nn  \\
&  + (1 - 2\epsilon )  \sum_{n=0}^{\infty} a_n (n + \delta) z^{n+\delta +1 }  + \sum_{n=0}^{\infty} a_n (n + \delta)(n + \delta -1) z^{n+\delta -1}   \nn\\
&  - l(l+1) \sum_{n=0}^{\infty} a_n  z^{n+\delta } - \mu^2 r_-^2 \sum_{n=0}^{\infty} a_n  z^{n+\delta }  -2 \sum_{n=0}^{\infty} a_n (n + \delta) (n + \delta -1) z^{n+\delta } \nn  \\
& + \sum_{n=0}^{\infty} a_n (n + \delta) (n + \delta -1) z^{n+\delta +1}  + 2i \Omega (r_+ - r_-) \sum_{n=0}^{\infty} a_n (n + \delta)  z^{n+\delta }  \nn\\
&  + \omega^2 r_-^2 \sum_{n=0}^{\infty} a_n z^{n+\delta +1} + \frac{2\omega r_-^2}{r_+ - r_-} (\omega r_+ - qQ) \sum_{n=0}^{\infty} a_n z^{n+\delta} \nn\\
& - \frac{2\omega r_-^2}{r_+ - r_-} (\omega r_+ - qQ) \sum_{n=0}^{\infty} a_n z^{n+\delta +1}  
\nn\\
& + \frac{ r_-^2}{{(r_+ - r_-)}^2} {(\omega r_+ - qQ)}^2 \sum_{n=0}^{\infty} a_n z^{n+\delta -1}
- 2 \frac{ r_-^2}{{(r_+ - r_-)}^2} {(\omega r_+ - qQ)}^2 \sum_{n=0}^{\infty} a_n z^{n+\delta } \nn\\
 & + \frac{ r_-^2}{{(r_+ - r_-)}^2} {(\omega r_+ - qQ)}^2 \sum_{n=0}^{\infty} a_n z^{n+\delta +1}  
\nn  \\
&+  4\omega r_- (\omega r_+ - qQ)  \sum_{n=0}^{\infty} a_n z^{n+\delta }  \nn\\
& +  \frac{ 2r_-}{r_+ - r_-} {(\omega r_+ - qQ)}^2 \sum_{n=0}^{\infty} a_n z^{n+\delta -1 } 
- \frac{ 2r_-}{r_+ - r_-} {(\omega r_+ - qQ)}^2 \sum_{n=0}^{\infty} a_n z^{n+\delta  } \nn
\end{align}
\begin{align}
& +  {(\omega r_+ - qQ)}^2 \sum_{n=0}^{\infty} a_n z^{n+\delta -1 }  \nn\\
& + \sum_{n=0}^{\infty} a_n z^{n+\delta  } \Big[ i\Omega (r_+ - r_-) + 2\omega (\omega r_+ - qQ)
(r_+ - r_-) - \mu^2 (r_+ + r_-) (r_+ - r_-)  \Big]   \nonumber\\
&  -iam qQ \frac{ (r_+ - r_-) \epsilon  (1-z)}{{(r_+ -z r_-)}^2} \sum_{n=0}^{\infty} a_n z^{n+\delta
 +1} \label{appendix2}\\
& - iam qQ (r_+ - r_-)   \frac{{(1-z)}^2}{{(r_+ -z r_-)}^2} \sum_{n=0}^{\infty} a_n (n+\delta) z^{n+\delta } \nn\\
& - iam qQ    \frac{{(1-z)}^2}{{(r_+ -z r_-)}^2} \Big(M - Q^2 \frac{1-z}{r_+ - zr_-}  \Big) \sum_{n=0}^{\infty} a_n z^{n+\delta }   \nn\\
& + amqQ \Omega    \frac{{(r_+ - r_-)}^2}{{(r_+ -z r_-)}^2} \sum_{n=0}^{\infty} a_n z^{n+\delta  +1}  =0, \nn
\end{align}
where $2M= r_+ + r_-$ and $Q^2 = r_+ r_-$.
Terms that do not involve the deformation parameter $a$ may combine \cite{QNMRNBrazilci, Chowdhury:2018izv} into an expansion leading to a $3$-term recurrence relation. In this way (\ref{appendix2}) is brought to the form
\begin{align}
& {(r_+ -z r_-)}^3 \sum_{n=1}^{\infty} \big[ \alpha_n a_{n+1} + \beta_n a_n + \gamma_n a_{n-1}   \big] z^{n+\delta}
+ {(r_+ -z r_-)}^3 \big(  \alpha_0 a_1 + \beta_0 a_0  \big) z^{\delta} \nn \\
&   + {(r_+ -z r_-)}^3 \Big[   \delta^2  +  {(\omega r_+ - qQ)}^2  \frac{r_+^2}{{(r_+ - r_-)}^2} \Big] a_0 z^{\delta -1} \nn \\
& - iamqQ (r_+ - r_-) \epsilon (r_+ - zr_-) (1-z)  \sum_{n=0}^{\infty} a_n z^{n+\delta  +1}   \label{appendix3}  \\
& - iamqQ (r_+ - r_-) (r_+ - zr_-) { (1-z)}^2  \sum_{n=0}^{\infty} a_n (n + \delta) z^{n+\delta } \nn\\
& - iamqQ (r_+ - zr_-) { (1-z)}^2  \sum_{n=0}^{\infty} a_n  z^{n+\delta }   \nn  \\
&  + iamqQ r_+ r_-  { (1-z)}^3  \sum_{n=0}^{\infty} a_n  z^{n+\delta } + am qQ \Omega {(r_+ - r_-)}^2 (r_+ -z r_-)  \sum_{n=0}^{\infty} a_n  z^{n+\delta + 1} = 0, \nn
\end{align}
with $\alpha_n, \beta_n, \gamma_n$  being written in (\ref{contfrsimple}). Since the general solution (\ref{generalpowersolution}) has to be nonsingular at the horizon $z=0$, and $z^{\delta -1} $ is a term of the lowest order in $z$, its coefficient in (\ref{appendix3})
has to vanish. This confirms the condition on $\delta$ that was obtained previously by matching with the QNM boundary conditions, see (\ref{epsilondelta}).

Next, we group together  terms with the same power of $z$
and shift the summation indices where needed to  bring all  terms into a form
having the same generic power $z^{n+\delta}$. Hence, we note that all   terms  determined
by $n \ge 4$  give rise to the general $6$-term
recurrence relation which characterizes the problem studied in this paper. 
Along with that, we need to sort out  all the remaining terms that do not fall into the above category. These terms give rise to the indicial recurrence relations. They are obtained by analysing the coefficients that respectively stand next to 
the powers $z^{\delta},z^{\delta +1},z^{\delta +2} $ and $z^{\delta +3}$. They may be considered as the boundary conditions imposed on the main, i.e. generic recurrence relation. The result is the  following  relation
\begin{align}
&  \sum_{n=4}^{\infty} \Big[  A_n a_{n+1} + B_n a_n +C_n a_{n-1} + D_n a_{n-2} + E_n a_{n-3} + F_n a_{n-4 } \Big] z^{n + \delta} \label{appendix4}   \\
&   + \bigg[  \Big( 3 r_+ r_-^2 \gamma_1 + iamqQ (r_+ - r_-) r_- \delta - r_-^3 \beta_0
- iamqQ (r_+ - r_-)r_- \epsilon + iamqQ M r_-      \nonumber   \\
& - iamqQ r_+ r_- \Big)a_0   +  \Big( 3 r_+ r_-^2 \beta_1 + iamqQ (r_+^2 - r_-^2)\epsilon  \nn\\ 
& - iamqQ (r_+ - r_-) (r_+ + 2r_-) ( \delta + 1)        \nonumber   \\   
&   -  iamqQ M (r_+ + 2r_-)   + 3iamqQ r_+ r_-  - amqQ\Omega {(r_+ - r_-)}^2 r_-   - 3 r_+^2 r_- \gamma_2 - r_-^3 \alpha_0 \Big) a_1   \nonumber\\
&    +  \Big(  r_+^3 \gamma_3 + iamqQ (r_+ - r_-)     (2r_+ - r_-)   ( \delta + 2) -3  r_+^2 r_- \beta_2
- iamqQ (r_+ - r_-) r_+ \epsilon       \nonumber   \\   
&   +  iamqQ M (2r_+  +r_-)   - 3iamqQ r_+ r_-  + amqQ\Omega {(r_+ - r_-)}^2 r_+   + 3  r_+ r_-^2   \alpha_1  \Big) a_2
\nonumber   \\
& +  \Big(  r_+^3 \beta_3 - iamqQ (r_+ - r_-) r_+   ( \delta + 3) -  iamqQ M  r_+    \nn\\
& + iamqQ r_+ r_-  -3  r_+^2   r_- \alpha_2  \Big) a_3 + r_+^3 \alpha_3  a_4 \bigg] 
z^{\delta+3} \nonumber  \\
&   + \bigg[  \Big( 3 r_+ r_-^2 \beta_0 + iamqQ (r_+ - r_-) (r_+ + r_-)  \epsilon   
- iamqQ (r_+ - r_-)(r_+ + 2r_-)  \delta  \nonumber   \\
& -  iamqQ M  (r_+ + 2 r_-)  +3 iamqQ  r_+ r_-  - amqQ \Omega  {(r_+ - r_-)}^2 r_-  - 3 r_+^2 r_- \gamma_1 \Big)a_0 \nonumber \\ 
&   +  \Big(  r_+^3 \gamma_2 + iamqQ (r_+ - r_-) (2r_+  + r_-)  (\delta +1)      - 3 r_+^2 r_- \beta_1
- iamqQ (r_+ - r_-)  r_+ \epsilon \nonumber   \\    
&  +  iamqQ M (2r_+ + r_-)   - 3iamqQ r_+ r_-  + amqQ\Omega {(r_+ - r_-)}^2 r_+   + 3 r_+ r_-^2 \alpha_0 \Big) a_1   \nonumber \\      
&  +  \Big(  r_+^3 \beta_2 - iamqQ (r_+ - r_-) r_+  ( \delta + 2) -   iamqQ M r_+   \nn\\
& -3  r_+^2 r_- \alpha_1 + iamqQ  r_+ r_-     \Big) a_2  + r_+^3 \alpha_2 a_3  \bigg] 
z^{\delta + 2} \nonumber   \\
&   + \bigg[  \Big(  r_+^3 \gamma_1   + amqQ \Omega {(r_+ - r_-)}^2 r_+ - 3 r_+^2 r_- \beta_0
- iamqQ (r_+ - r_-) r_+  \epsilon   \nn\\
& + iamqQ (r_+ - r_-)(2r_+ + r_-)\delta +  iamqQ M  (2r_+ +  r_-)  - 3 iamqQ  r_+ r_-     \Big)a_0 
\nonumber\\ 
& +  \Big(  r_+^3 \beta_1 - iamqQ (r_+ - r_-) r_+   (\delta +1)  -  iamqQ M r_+       \nonumber \\  
& +   iamqQ r_+  r_-    - 3 r_+^2 r_- \alpha_0  \Big) a_1  + r_+^3 \alpha_1 a_2   \bigg] 
z^{\delta +1}    \nonumber
\end{align}
\begin{align}
&  +   \bigg[ \Big(  r_+^3 \beta_0 + iamqQ r_+   r_-   - iamqQ (r_+ - r_-)r_+ \delta   -   iamqQ M
r_+  \Big)a_0  + r_+^3 \alpha_0 a_1  \bigg]  z^{\delta } =0. \nn
\end{align}
The coefficients $A_n, B_n, C_n, D_n, E_n, F_n$ that appear here are given in (\ref{contfr1}). Moreover, it is easy to see that the expressions multiplying $a_0, a_1, a_2, a_3, a_4$ that appear in front of the lowest order powers 
$z^{\delta},z^{\delta +1},z^{\delta +2} $ and $z^{\delta +3}$ may be recognized as the particular cases of the coefficients
$A_n, B_n, C_n, D_n, E_n, F_n$. Indeed, these expressions may be written in terms of $A_i, B_i, C_i, D_i, E_i, ~ i=0,1,2,3$.
Thus, the sequence of recurrence relations following from (\ref{appendix4}) can be written as
\begin{eqnarray}  
A_n a_{n+1} + B_n a_n +C_n a_{n-1} + D_n a_{n-2} + E_n a_{n-3} + F_n a_{n-4 }  = 0, \nonumber \\
A_3 a_{4} + B_3 a_3 +C_3 a_{2} + D_3 a_{1} + E_3 a_{0}   = 0, \nonumber \\
A_2 a_{3} + B_2 a_2 +C_2 a_{1} + D_2 a_{0}   = 0, \nonumber \\
A_1 a_{2} + B_1 a_1 +C_1 a_{0}    = 0, \nonumber \\
A_0 a_{1} + B_0 a_0   = 0. \label{App6TermFinal}
\end{eqnarray}

\section{Gauss elimination procedure}

In the following we explain the required steps in a Gauss elimination procedure applied to our particular model. We need to reduce the $6$-term recurence relation (\ref{6contfr}) to a $3$-term recurence relation.

To begin with, we define the coefficients of the zeroth level $A_n^{(0)}, B_n^{(0)}, C_n^{(0)}, D_n^{(0)},$ $E_n^{(0)}, F_n^{(0)}$ to be the coefficients  of the initial $6$-term recurrence relation (\ref{6contfr}),
\begin{equation} \label{zerothorder}
 A_n^{(0)} \equiv A_n, ~  B_n^{(0)} \equiv B_n, ~  C_n^{(0)} \equiv C_n, ~ D_n^{(0)} \equiv D_n, ~  E_n^{(0)} \equiv E_n, ~  F_n^{(0)} \equiv F_n,
\end{equation}
with $A_n, B_n, C_n, D_n, E_n, F_n$ defined in (\ref{contfr1}).
Similarly, we introduce the coefficients of the $jth$ level $A_n^{(j)}, B_n^{(j)}, C_n^{(j)}, D_n^{(j)}, E_n^{(j)}$, $j=1,2,3$ as the coefficients that appear in the $(6-j)$-term recurence relation, obtained after the $jth$ Gauss elimination.

Applying the first Gauss elimination to (\ref{6contfr})  we find  the $5$-term recurrence relation
\begin{eqnarray}  \label{contfr5term}
  A_n^{(1)} a_{n+1} + B_n^{(1)} a_n +C_n^{(1)} a_{n-1} + D_n^{(1)} a_{n-2} + E_n^{(1)} a_{n-3}   &=& 0, \nonumber \\
  A_2^{(1)} a_{3} + B_2^{(1)} a_2 +C_2^{(1)} a_{1} + D_2^{(1)} a_{0}   &=& 0, \nonumber \\
  A_1^{(1)} a_{2} + B_1^{(1)} a_1 +C_1^{(1)} a_{0}    &=& 0, \nonumber \\
   A_0^{(1)} a_{1} + B_0^{(1)} a_0   &=& 0. 
\end{eqnarray}
The coefficients of the first level are determined as 
\begin{align}    \label{contfr5termcoeff1}
 A_n^{(1)} = A_n^{(0)},  \qquad \qquad \qquad &&     n\ge 4       &&      \nonumber \\
   B_n^{(1)}  = B_n^{(0)} - \frac{F_n^{(0)}}{E_{n-1}^{(1)}} A_{n-1}^{(1)},  && C_n^{(1)}  =   C_n^{(0)} - \frac{F_n^{(0)}}{E_{n-1}^{(1)}} B_{n-1}^{(1)},   &&       \nonumber
\\
  D_n^{(1)} =     D_n^{(0)} - \frac{F_n^{(0)}}{E_{n-1}^{(1)}} C_{n-1}^{(1)}, &&  E_n^{(1)} =  E_n^{(0)} - \frac{F_n^{(0)}}{E_{n-1}^{(1)}} D_{n-1}^{(1)},   &&   
\end{align}
and for $n = 3,2,1,0$
\begin{align}    \label{contfr5termcoeff}
   A_3^{(1)} = A_3^{(0)},   &&   B_3^{(1)} = B_3^{(0)},      &&  C_3^{(1)} = C_3^{(0)},  &&   D_3^{(1)} = D_3^{(0)},  &&  E_3^{(1)} = E_3^{(0)},       \nonumber \\
A_2^{(1)} = A_2^{(0)},     &&   B_2^{(1)} = B_2^{(0)},      &&  C_2^{(1)} = C_2^{(0)}, && D_2^{(1)} = D_2^{(0)},      \nonumber \\
  A_1^{(1)} = A_1^{(0)},  &&  B_1^{(1)} = B_1^{(0)},        && C_1^{(1)} = C_1^{(0)},  &&     &&    \nonumber  \\
  A_0^{(1)} = A_0^{(0)},   &&  B_0^{(1)} = B_0^{(0)}. &&   &&   &&   
\end{align}
The application of the second Gauss elimination to the recurrence relation (\ref{contfr5term}) yields the following $4$-term recurrence equation
\begin{eqnarray}  \label{contfr4term}
  A_n^{(2)} a_{n+1} + B_n^{(2)} a_n +C_n^{(2)} a_{n-1} + D_n^{(2)} a_{n-2}   = 0, \nonumber \\
  A_1^{(2)} a_{2} + B_1^{(2)} a_1 +C_1^{(2)} a_{0}    = 0, \nonumber \\
   A_0^{(2)} a_{1} + B_0^{(2)} a_0   = 0. 
\end{eqnarray}
The coefficients of the second level are determined as
\begin{align}    \label{contfr4termcoeff}
  A_n^{(2)} = A_n^{(1)} = A_n^{(0)},   \quad \qquad &&   B_n^{(2)} = B_n^{(1)} - \frac{E_n^{(1)}}{D_{n-1}^{(2)}} A_{n-1}^{(2)},          &&   n\ge 3    \nonumber \\
   C_n^{(2)}  =   C_n^{(1)} - \frac{E_n^{(1)}}{D_{n-1}^{(2)}} B_{n-1}^{(2)},    &&    D_n^{(2)} =     D_n^{(1)} - \frac{E_n^{(1)}}{D_{n-1}^{(2)}} C_{n-1}^{(2)},     &&    
\end{align}
and for $n = 0,1,2,$
\begin{align}     \label{contfr4termcoeff1}
   A_2^{(2)} = A_2^{(1)} = A_2^{(0)},   &&    B_2^{(2)} = B_2^{(1)} = B_2^{(0)},     &&   C_2^{(2)} = C_2^{(1)} = C_2^{(0)},  &&    D_2^{(2)} = D_2^{(1)} = D_2^{(0)},    \nonumber \\
  A_1^{(2)} = A_1^{(1)}= A_1^{(0)},    &&   B_1^{(2)} = B_1^{(1)} = B_1^{(0)},     &&   C_1^{(2)} = C_1^{(1)} = C_1^{(0)},
 &&     \nonumber   \\
  A_0^{(2)} = A_0^{(1)} =  A_0^{(0)},  &&  B_0^{(2)} = B_0^{(1)} = B_0^{(0)}.     &&   &&     
\end{align}
Third and the last Gauss elimination applied to (\ref{contfr4term}) leads to the recurrence relation
\begin{eqnarray}    \label{contfrsimplegauss}
A_n^{(3)} a_{n+1} + B_n^{(3)} a_n +  C_n^{(3)} a_{n-1} &=& 0, \nonumber \\
A_0^{(3)} a_{1} + B_0^{(3)} a_0   &=& 0.  
\end{eqnarray}
The  coefficients of the third level,  $A_n^{(3)}, B_n^{(3)}, C_n^{(3)}, $ that we have been searching for, are given by
\begin{align}
A_n^{(3)} = A_n^{(2)} = A_n^{(0)},   \quad \qquad &&  n\ge 2    \nonumber \\
B_n^{(3)} = B_n^{(2)} - \frac{D_n^{(2)}}{C_{n-1}^{(3)}} A_{n-1}^{(3)}, 
&& C_n^{(3)}  =   C_n^{(2)} - \frac{D_n^{(2)}}{C_{n-1}^{(3)}} B_{n-1}^{(3)},  \label{contfr3termcoeff}
\end{align}
and for $n = 0,1,$
\begin{align}     \label{contfr3termcoeff1}
A_1^{(3)} = A_1^{(2)}= A_1^{(0)}, &&   B_1^{(3)} = B_1^{(2)} = B_1^{(0)},       &&  C_1^{(3)} = C_1^{(2)} = C_1^{(0)},   \nonumber   \\
A_0^{(3)} = A_0^{(2)} =  A_0^{(0)},   &&    B_0^{(3)} = B_0^{(2)} = B_0^{(0)}.    && 
\end{align}


\end{document}